# Design Principles for Heterointerfacial Alloying Kinetics at Metallic Anodes in Rechargeable Batteries

Jingxu Zheng[1,2][†], Yue Deng[1][†], Wenzao Li[3,4], Jiefu Yin[5], Patrick J. West[4,6], Tian Tang[1], Xiao Tong[7], David C. Bock[4,8], Shuo Jin[5], Qing Zhao[5], Regina Garcia-Mendez[5], Kenneth J. Takeuchi[3,4,6,8], Esther S. Takeuchi[3,4,6,8], Amy C. Marschilok[3,4,6,8], Lynden A. Archer[1,5*]

1. Department of Materials Science and Engineering, Cornell University, Ithaca, NY 14853, USA.

2. Department of Physics, Massachusetts Institute of Technology, Cambridge, MA 02139, USA.

3. Department of Chemistry, State University of New York at Stony Brook, Stony Brook, NY 11794, USA.

4. Institute for Electrochemically Stored Energy, Stony Brook University, Stony Brook, NY 11794, USA.

5. Robert Frederick Smith School of Chemical and Biomolecular Engineering, Cornell University, Ithaca, NY 14853, USA.

6. Department of Materials Science and Chemical Engineering, State University of New York at Stony Brook, Stony Brook, NY 11794, USA.

7. Center for Functional Nanomaterials, Brookhaven National Laboratory, Upton, NY 11973, USA.

8. Interdisciplinary Science Department, Brookhaven National Laboratory, Upton, NY 11973, USA.

*\* Corresponding author: laa25@cornell.edu*

[†]These authors contributed equally to this work.

**Abstract:** How surface chemistry influences reactions occurring thereupon has been a long-standing question of broad scientific and technological interest. Here we consider the relation between the surface chemistry at such interfaces and the reversibility of electrochemical transformations at rechargeable battery electrodes. Using Zn as a model system, we report that a moderate strength of chemical interaction between the deposit and the substrate—neither too weak nor too strong—enables highest reversibility and stability of the plating/stripping redox processes. Focused-ion-beam and electron microscopy were used to directly probe the morphology, chemistry and crystallography of the heterointerfaces of distinct natures. Analogous to the empirical *Sabatier* principle for chemical heterogeneous catalysis, our finding arises from the confluence of competing interfacial processes. Using full batteries with stringent N:P ratios, we show that such knowledge provides a powerful tool for designing key materials in highly reversible battery systems based on earth-abundant, low-cost metals such as Zn and Na.

**One sentence summary:** Principles of interfacial chemical kinetics provide powerful guidelines for designing battery anodes.

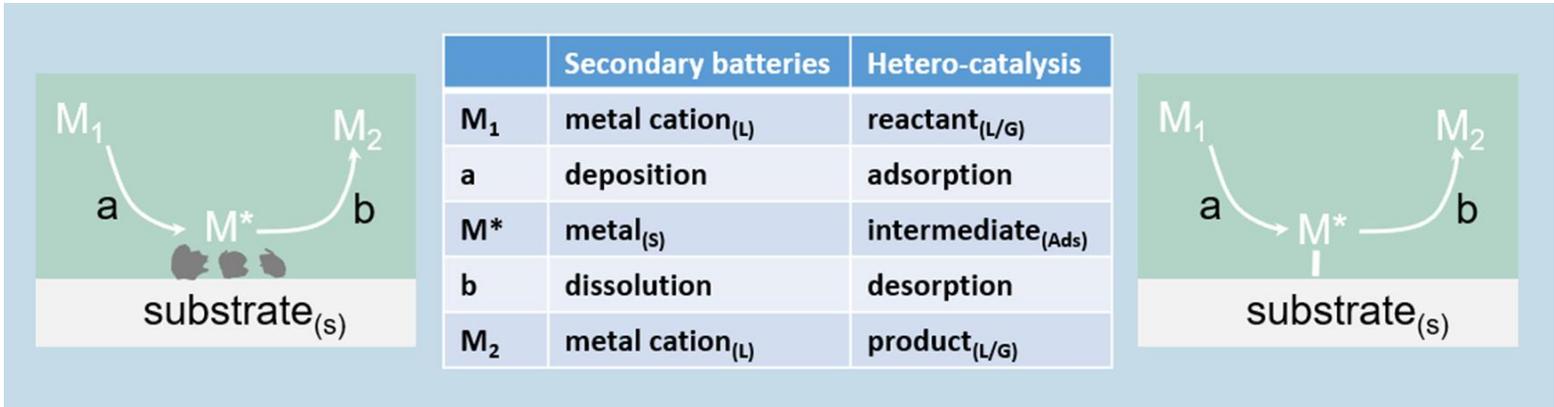

**Figure 1. Reaction path at a heterointerface: in secondary batteries (left), in heterogeneous catalysis (right).** In both scenarios, the reactant species first "attaches" to and then "detaches" from the surface of a solid, as labelled by "a" and "b", respectively.

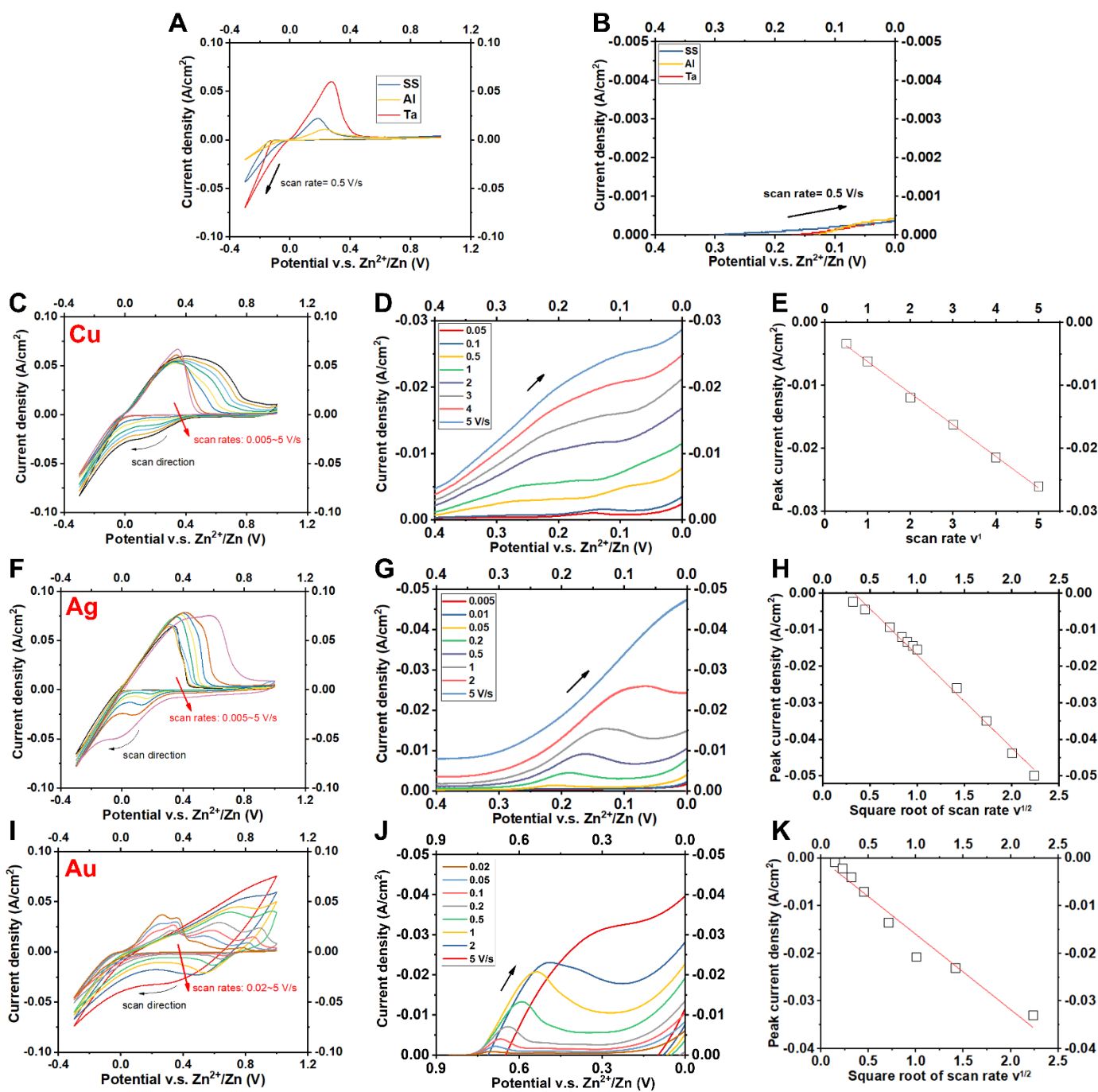

**Figure 2. Cyclic voltammetry (CV) of Zn metal plating/stripping on substrates of different chemistries.** CV scans in the range -0.3 to +1.0 V v.s. $Zn^{2+}/Zn$ at various metal electrodes substrates in a conventional 2 M $ZnSO_4$ (aq) electrolyte: (A) stainless steel, Al and Ta, respectively; (C) Cu; (F) Ag; (I) Au. (B)(D)(G)(J) Enlarged plots showing the underpotential regime in a negative scan of the substrates in (A), (C), (F), and (I), respectively. The scan rates in (C), (F) and (I) progress in the same way as in (D), (G) and (J). (E) Plot of peak current density versus scan rate ($i_p$ versus $v$ in V/s) for Cu. Plots of peak current density versus square root of scan rate ($i_p$ versus $v^{1/2}$, in V$^{1/2}$/s$^{1/2}$) for (H) Ag and (K) Au.

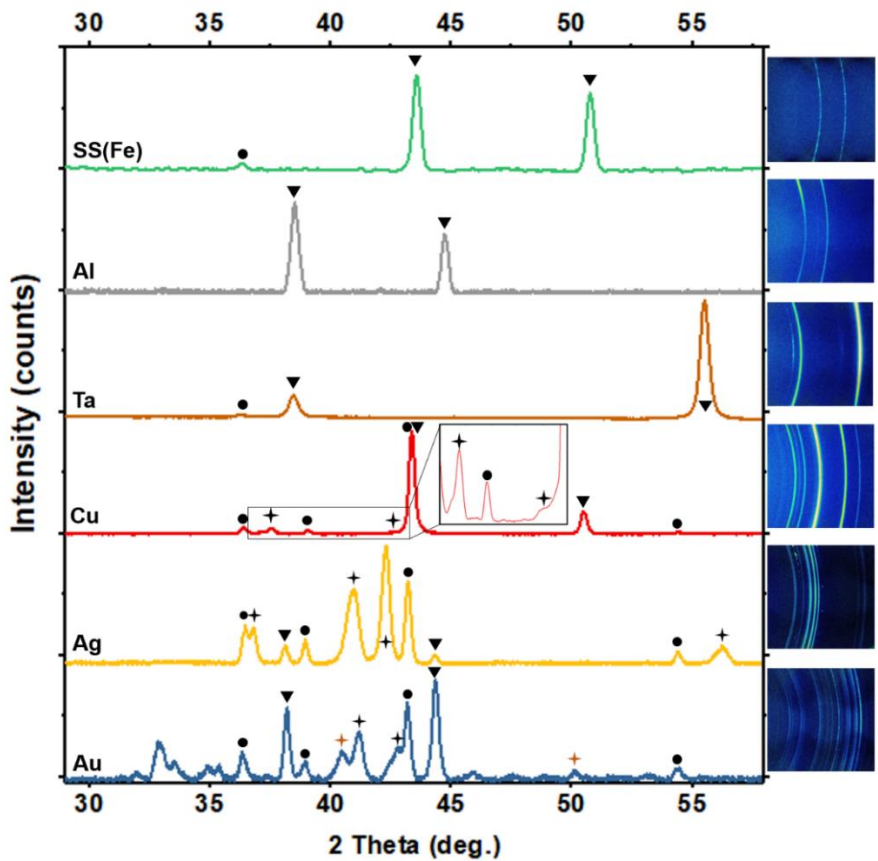

**Figure 3. X-ray diffraction analysis of various metal substrates after Zn plating/stripping cycling.** Notation: ●Zn metal; ▼Substrates; ┼ Intermetallic phases. The respective 2D XRD patterns are provided on the right. Sample condition: after 100 cycles at 8 mA/cm$^2$, 0.8 mAh/cm$^2$; with 0.8 mAh/cm$^2$ Zn deposition remaining on the substrate before cell dissembling. Zn deposits could be peeled off and stuck in the porous glass-fiber separators if the interface between Zn and the substrates is not mechanically robust. As such, the Zn deposits may not appear in the XRD patterns.

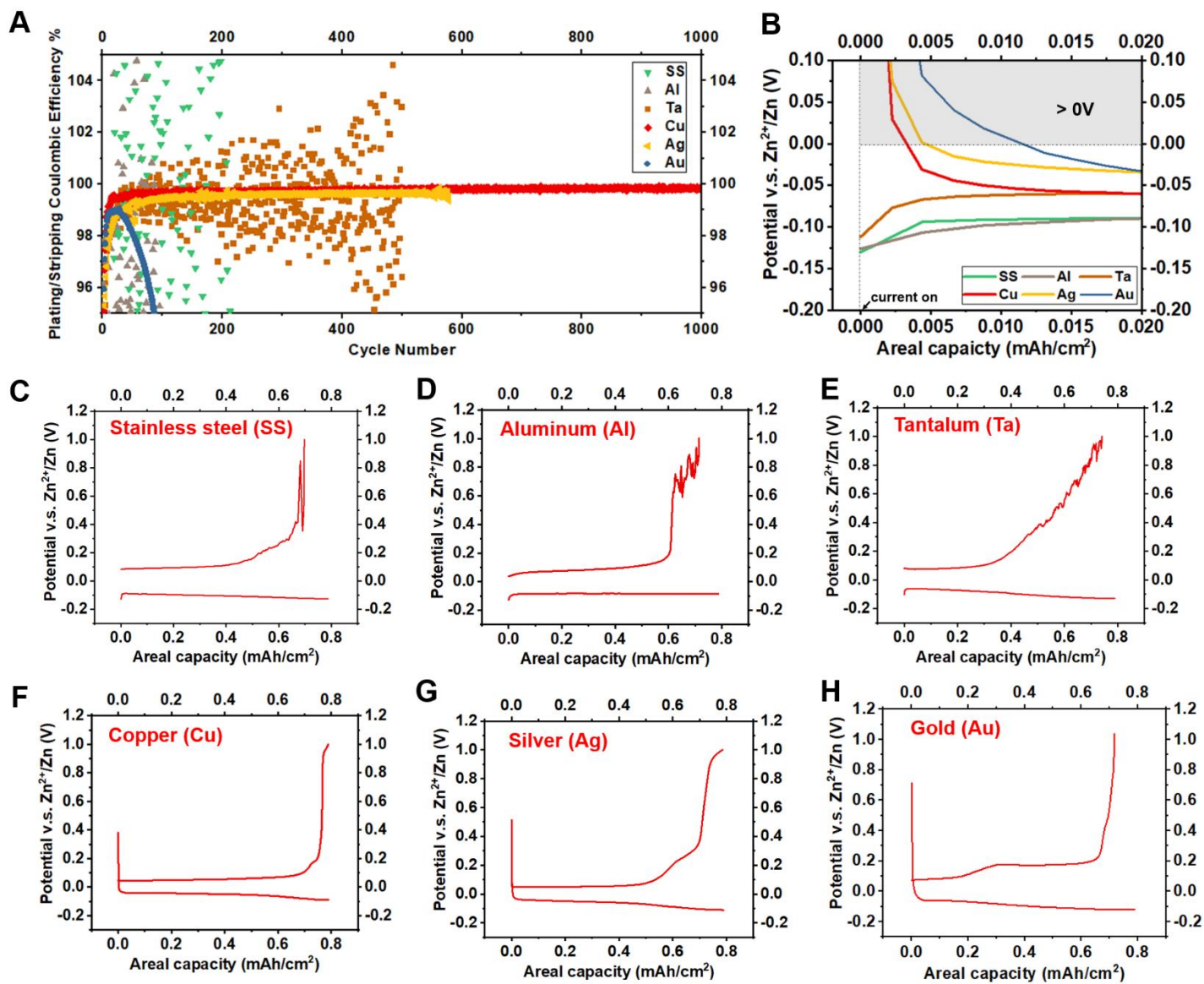

**Figure 4. Galvanostatic electrochemical deposition/dissolution behaviors of Zn on substrates of different chemistry.** (A) Coulombic efficiency for Zn plating/stripping as a function of cycle number. (B) Voltage evolution in the nucleation regime. Representative plating/stripping voltage profiles of Zn plating/stripping on (C) stainless steel, (D) Al, (E) Ta, (F) Cu, (G) Ag, and (H) Au. The spiky nature of the stripping voltage profiles observed on SS, Al and Ta is consistent with expectations for formation of "orphaned" Zn metal. Secondary voltage plateaus are observed on Cu, Ag and Au. Current density: 8 mA/cm$^2$.

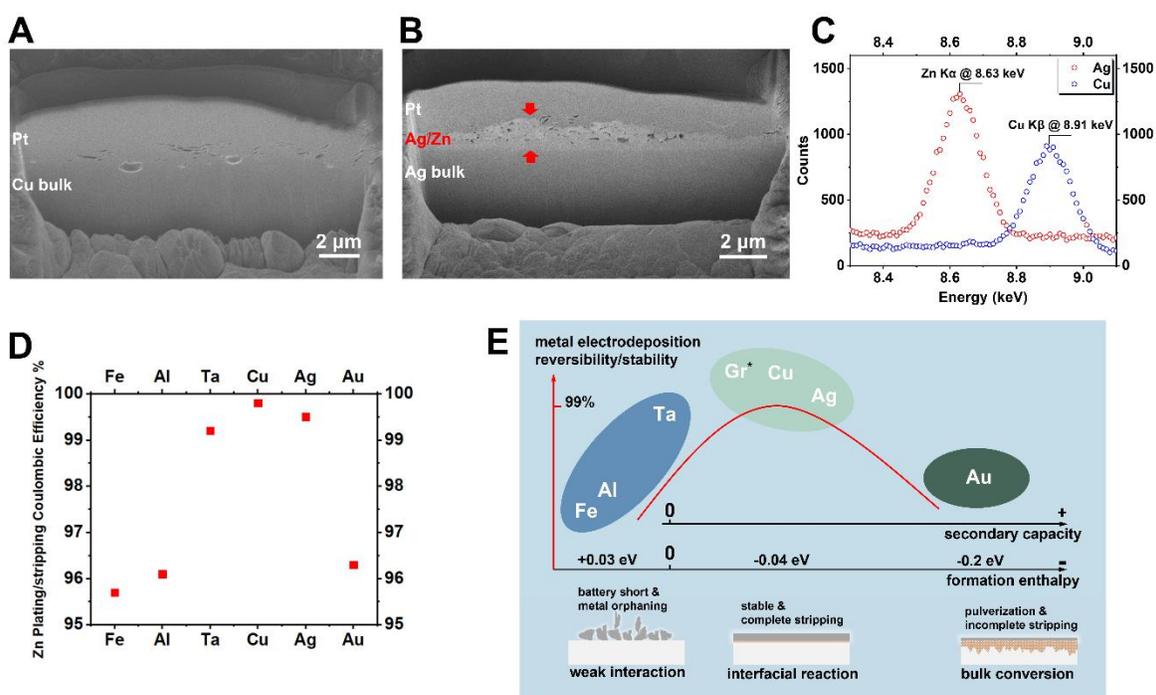

**Figure 5. A Sabatier-like principle for reversible metal deposition/dissolution at battery anodes.** FIB-SEM characterization of (A) Cu and (B) Ag electrodes after Zn plating/stripping, respectively. A Pt protection layer is deposited on electrode surface before FIB. (C) EDS near the Zn Kα energy (*i.e.*, 8.63 keV) collected at the two interfaces formed on Ag and Cu, respectively. See more details in **Fig. S12~S15**. (D) The average values and the standard deviations of the metal plating/stripping Coulombic efficiency of the substrates. (E) The proposed qualitative volcano-shaped relation in battery anodes. Gr* stands for graphene.

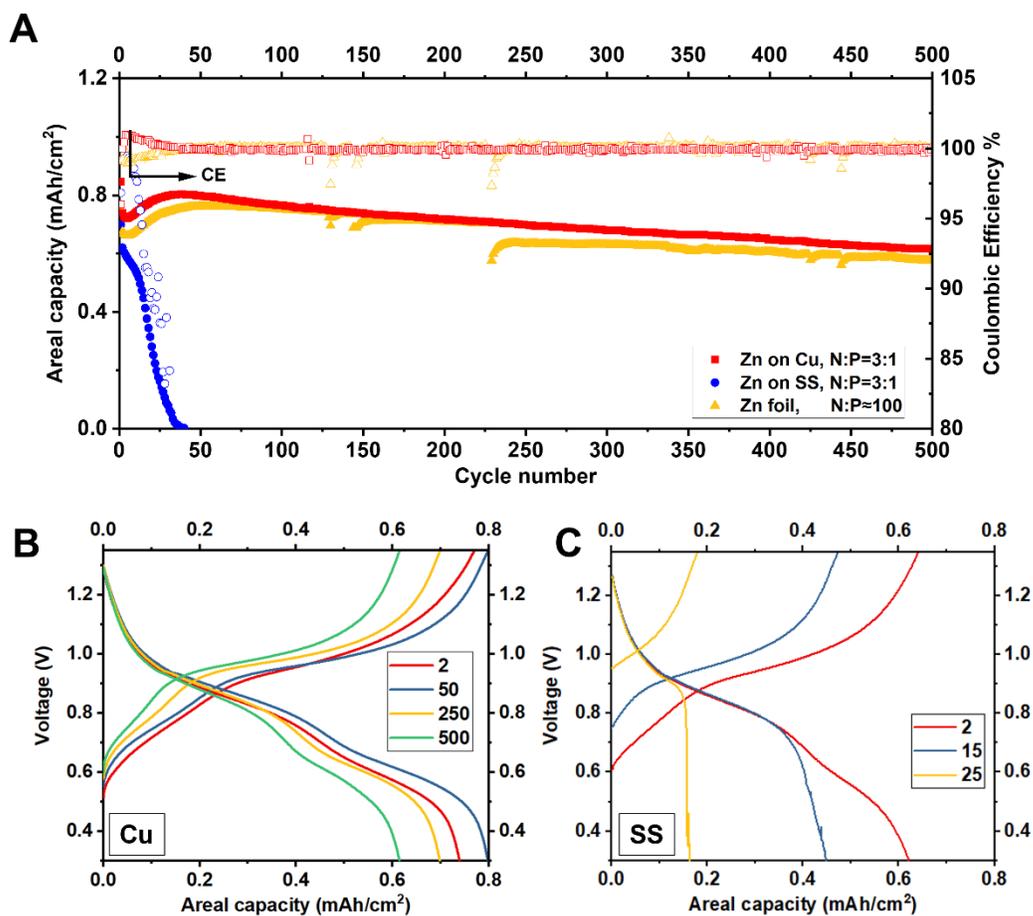

**Figure 6. Full battery cell demonstration under stringent active material loading conditions.** (A) Cycling performance of Zn||NaV$_3$O$_8$ full cells. A certain amount of Zn metal is deposited on Cu or stainless steel (SS) and used as the anode. The performance of commercial thick Zn foil is also tested. The areal capacity is ~0.7 mAh/cm$^2$, the current density is 1.3 mA/cm$^2$, and the cathode area is ~1 cm$^2$; the negative electrode to positive electrode capacity (N:P) ratio is maintained at 3:1 for Zn on Cu and Zn on SS. Charge-discharge voltage profiles of batteries using Zn deposited on (B) Cu and (C) stainless steel, respectively. The numbers in the legends of (B) and (C) denote the cycle numbers. Solid data points and open data points represent the capacity (left y axis) and the Coulombic efficiency (right y axis), respectively.

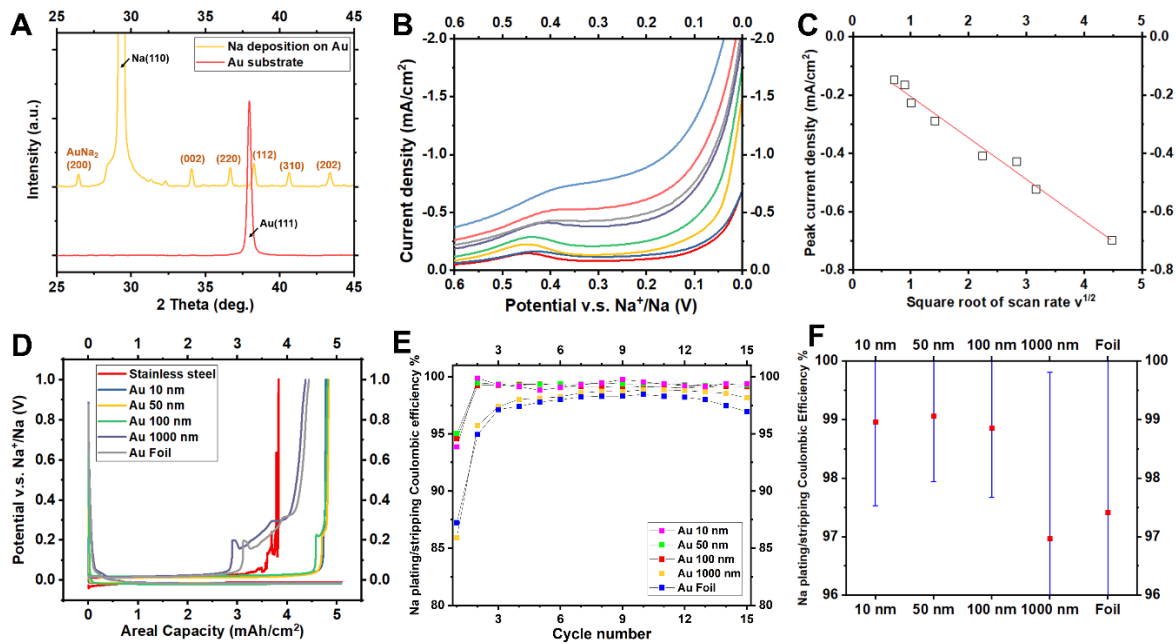

**Figure 7. Evaluation of the Sabatier-like principle for Na metal battery anodes.** (A) XRD of the original Au substrate (red) and the Au substrate after Na electrodeposition (yellow). Substrate: 100 nm Au deposited on stainless steel. (B) CV scans of Na on Au foil at representative rates. (C) Plot of peak current density versus the square root of scan rate ($i_p$ versus $v^{1/2}$, in $V^{1/2}/s^{1/2}$). (D) Galvanostatic plating/stripping voltage profiles for Na on Au substrates of different thicknesses (*i.e.*, bare stainless steel, 10 nm Au, 50 nm Au, 100 nm Au, 1000 nm Au, and thick foil-type Au). Galvanostatic plating/stripping Coulombic efficiency values for Na on Au substates with different thicknesses: (E) distribution over cycle number; and (F) run-to-run averages and statistics.

**Introduction**

Control of chemical reaction kinetics at a heterointerface is required in multiple chemical transformations, including ammonia synthesis (*1*), hydrocarbon reforming (*2*), water gas shift reactions (*3*), *etc.* in order to achieve economic viability at commercial scale. All of these processes typically involve multiple and complex physiochemical kinetic steps occurring on the surface of a solid with specific catalytic activity for one or more reactants; they are here collectively termed heterocatalysis. Driven by the fundamental challenge of rationally designing substrates with high and selective activity to speed up conversion of inexpensive reactants into high-value products, and commercial interest in performing these reactions on a large scale, research centering on heterointerfacial chemical reactions has flourished for at least the last one hundred years (*4, 5*). Here we consider a successful empirical design concept, the ***Sabatier principle***—that has provided a useful framework for organizing heterocatalysis data—in what might at first appear an unrelated context, electrical energy storage in batteries using metallic anodes. The *Sabatier* principle states simply that the highest chemical activity is achieved when the interaction between reactant molecules and the catalyst surface is at an optimal value (*6, 7*), *i.e.*, neither too weak nor too strong.

**Figure 1** is the starting point of our analysis. It illustrates the similarities between the physicochemical sequence that leads to growth/dissolution of a metal electrode in a rechargeable battery and those that govern reversible chemical transformations on the surface of catalyst particles, *e.g.*, hydrogen evolution reaction and hydrogen oxidation reaction on Pt (*8*). Considering that unlike the latter scenario, the reactant undergoes a coupled redox reaction and phase change in the former, we wondered whether an analogous Sabatier-like design principle exists, and whether the principle could be used for electrode design/selection in rechargeable metal batteries to achieve high reversibility. The answer to this question is particularly critical to metal anodes with low N:P ratios or using a so-called "anode-free" configuration; in these cases, reversible metal plating/stripping on a heterosubstrate—as opposed to the metal itself—is almost inevitable over cycling.

As the first step to evaluate this conjecture, we investigated zinc metal plating/stripping in aqueous electrolytes. Our motivation for studying aqueous zinc batteries is straightforward. First, secondary/rechargeable batteries based on earth-abundant and low-cost metals such as Zn, Na and Al, are among the most promising solutions for high-energy-density, low-cost, and long-term storage of electric power at scales compatible with the rapidly rising scale of global production of electricity from renewable wind and solar photovoltaic resources (*9, 10*), and which can handle both the diurnal and seasonal variability in energy supply from these sources (*11, 12*). Second, Zn metal has a moderate redox potential, which means that electrochemical processes at a Zn electrode can be more easily decoupled from chemical and electrochemical parasitic reactions with electrolyte components, which are known to lower the reversibility of metals such as Li and Na (*10*), allowing us to focus on the chemical kinetics between the metal and the substrate at the heterointerface. Finally, analogous to conventional heterocatalytic processes (*13*), the electrochemical and physical transformations that control the charge (electroreduction) and discharge (ion dissolution/ solvation) processes at the solid-liquid heterointerface in a Zn battery electrode, place quite different and even competing demands for the electrode and electrolyte design.

It is also known that in order for rechargeable batteries of any chemistry to offer cost-competitive storage of electric power, high levels of electrode reversibility are required to enable long operating lifetimes and, consequently, low amortized system costs. Understanding and controlling the chemical kinetics of reversible deposition/dissolution of metals at heterointerfaces in battery electrodes are therefore fundamental requirements for progress. On the one hand, it is understood that the irreversibility of plating/stripping processes at a metal battery anode is the dominant source of degradation of storage capacity with time (*9*) and is also associated with serious safety risks associated with internal shorting (*14*). On the other hand, multiple physiochemical processes are actively involved, such as crystal nucleation and growth (*15*), mass diffusion, hydrodynamic convection, *etc*.—and therefore control is challenging (*16, 17*). A diversity of strategies have been proposed to address these challenges, including electrolyte design (*18, 19*), electrode architecture design (*20, 21*), and surface chemistry/interphase design (*12, 22*). Redox chemistry at heterointerfaces has also been leveraged in some studies, *via* intentional, *apriori* chemical design of the substrate or through ion-exchange chemistry of electrolyte additives to regulate growth of the electrodeposited metals. Strong chemical binding/interactions between the substrate and the metal deposited at the heterointerface—also known as the "metal-philicity" —is conventionally thought to improve control of deposit morphology in either case, by enabling a uniform nucleation landscape for electrodeposit growth (*23, 24*).

A first requirement for our studies is to create a series of heterointerfaces that are of high quality, exhibit good uniformity, and which enable the interaction strengths with Zn metal to be tuned. A common approach for manipulating binding strength is to dope carbonaceous materials with heteroatoms (*22*). The approach is unattractive here because it introduces complications (*e.g.*, the typically used carbon-based materials are associated with non-trivial geometries, *e.g.* anisotropic shape, high specific surface area, *etc*.) (*25*), which could play additional unexpected roles in electrodeposition (*e.g.*, by altering the local electric field distribution at the substrate) (*12*);

confounding any serious efforts at deconvoluting and evaluating the contribution from interfacial chemical kinetics.

Here, we instead leverage Zn's known capability to form alloys with other metals upon electro-reduction as reported in early literature (*26, 27*). These metals are commercially available as foils, which offer a well-defined, planar geometry as is evident from SEM topographic and AFM roughness analysis (**Fig. S1**). Given the rich chemistries achievable on metals that have different interaction strengths with Zn, it is possible to semi-quantitatively evaluate the role played by heterointerfacial chemical kinetics in electroplating/stripping.

**Results**

We first utilized cyclic voltammetry (**CV**) as a classical electroanalytical tool (*28*) to interrogate the electrochemical reactions occurring at a representative set of metal substrates, in a range of chemistries: stainless steel (**SS**), aluminum (**Al**), tantalum (**Ta**), copper (**Cu**), silver (**Ag**) and gold (**Au**). In the CV experiments, the metal substrate is used as the working electrode in an electrochemical cell and is immersed in a conventional Zn battery electrolyte— 2 M $ZnSO_4$ (aq). The results reported in **Fig. 2** show the current-potential (*i*-V) responses. The differences are stark, large, and systematic. On SS, Al and Ta, only the metal plating and stripping peaks are detectable. For example, in a negative scan, the response currents above 0 V v.s. $Zn^{2+}$/Zn are negligible, *i.e.*, < 0.5 mA/cm$^2$ (**Fig. 2A~B**), and the standard $Zn^{2+} \rightarrow Zn_{(s)}$ reaction is detected as the potential enters the regime below 0 V v.s. $Zn^{2+}$/Zn. In contrast, response currents two orders of magnitude higher are observed above 0 V v.s. $Zn^{2+}$/Zn on Cu (**Fig. 2C~E**), Ag (**Fig. 2F~H**) and Au (**Fig. 2I~K**), respectively, in addition to the Zn plating current below 0 V v.s. $Zn^{2+}$/Zn.

Clues about the origins of the $i$-V responses can be found in results reported in prior studies. Specifically, the "underpotential deposition" behaviors of Zn on Cu, Ag and Au are attributable to spontaneous alloying between the metal substrate and metal deposits (*29*). The Zn alloys exhibit a lower free energy than elemental Zn, resulting in a smaller potential required to reduce $Zn^{2+}$ (*30*). Further quantitative examination of the $i$-V responses of Cu, Ag and Au reveals subtle but critical differences in alloying mechanism. We plot the peak current $i_p$ over a series of scan rates $v$ in **Fig. 2E**, **H** and **K**. Ag and Au show a characteristic $i_p \sim v^{1/2}$ scaling behavior, while Cu manifests a linear relation $i_p \sim v^1$. The $R^2$ values of the fittings are 0.99, 0.96 and 0.998 for Ag, Au and Cu, respectively. We conclude that whereas Ag and Au undergo diffusion-controlled, bulk-conversion-type alloying processes as described by the Randles-Sevcik equation, alloying between Cu-Zn progresses by a "pseudocapacitive"-like route (*31*). This suggests that the Cu-Zn alloying reaction is confined to a thin skin layer on the substrate, which is consistent with the pseudo-capacitive $i_p \sim v$ scaling relation apparent from the CV experiments.

Conclusions from our indirect electroanalytical analyses of the reaction currents are corroborated by findings from more in-depth characterization of the electrodeposited phases. X-ray diffraction is used to directly capture the crystallographic transitions occurring on the substrates (**Fig. 3**). The results show that SS, Al and Ta remain chemically intact after Zn plating, not undergoing any detectable reaction with Zn metal. In contrast, diffraction patterns assigned to intermetallic compounds, *i.e.*, $\varepsilon$-$CuZn_4$, $\varepsilon$-$AgZn_3$, $AuZn_3$ + $AuZn$, are detected on Cu, Ag, Au, respectively. The significantly stronger alloy peaks (labelled by the cross ✣) observed for Ag and Au, relative to those seen for the Cu substrates, implies that the amounts of alloy formed on the former two are substantially greater than the latter after cycling. This observation is consistent with the argument that Ag and Au undergo bulk conversion with Zn, but the reaction on Cu is limited to a thin layer.

**Figure 4** reports the electrochemical galvanostatic plating/stripping behaviors of Zn on the same metal foil substrates. As can be clearly seen in **Fig. 4A**, the substrate chemistry has a strong influence on the plating/stripping cycling reversibility of Zn as quantified by the Coulombic efficiency (CE) = $\frac{\text{stripping capacity}}{\text{plating capacity}} \times 100\%$ (see also **Methods** and **Fig. S2** for additional information). We note further that the current density used is well below the diffusion limit (*32*). These results therefore lead to two conclusions: (i) SS, Al and Ta are less effective substrates for achieving high Zn reversibility because they lead to a significantly broader distribution of CE values than Cu, Ag and Au, indicating unstable plating/stripping process; (ii) Cu, Ag and Au all show relatively narrow CE distributions, but Au has a significantly lower CE value corresponding to low plating/stripping reversibility.

The diverse plating/stripping behaviors on different metal substrates can be understood by examining the voltage profiles of the plating/stripping cycles as shown in **Fig. 4B~H**. For SS, Al and Ta where broad CE distributions are seen, the stripping process consistently features spiky, erratic voltage evolution; in stark contrast, the voltage profiles of Zn stripping on Cu, Ag and Au are smooth and stable (see also **Fig. S3** for Zn plating/stripping on Cu with a lower cutoff voltage). The voltage profiles of SS, Al and Ta are indeed suggestive of an undesirable process termed *metal orphaning* (*33-35*). The behavior of Ti foil, another conventional substrate used for evaluating Zn plating/stripping, is similar to SS, Al and Ta (**Fig. S4**). See also **Fig. S5** and **Fig. S6** for measurements made at other capacities. In the orphaning process, the metallic, solid electrodeposits are physically disconnected from the electron source at the electrode, thereby becoming electrochemically inactive/dead. The sudden potential drops and the sometimes >100% Coulombic efficiency values are caused by random re-connection of inactive/dead metal fragments over cycling (*32, 35*). Both the non-uniformity of the deposition and ease of physical detachment from the substrate are related to the propensity of the electrodeposited metal to orphan.

To experimentally verify this interpretation, we use optical microscopy (OM) and scanning electron microscopy (SEM) to characterize the separator membrane facing the heterointerface after plating/stripping cycling (**Fig. S7~8**). The results show that large amounts of Zn metal electrodeposits on SS, Al and Ta become electrochemically inactive and are stuck in the separator during repeated plating/stripping, while no dead Zn is observable on Cu and Ag substrates. On the separator facing an Au substrate, coarse, dark brownish powders are observed under the optical microscope, which are identified as an intermetallic phase (AuZn, brown cross ✚) by X-ray diffraction (**Fig. S9**), as opposed to dead metallic Zn.

We further examined the galvanostatic plating/stripping voltage profiles to assess the resistances to electrochemical change. The first observation is that a prominent nucleation overpotential exists at the onset of Zn plating on SS, Ta and Al, but not on Cu, Ag and Au; instead, significant capacities above 0 V v.s. $Zn^{2+}$/Zn are detected that are attributable to the alloying reactions. Spontaneous alloying processes can also occur after the electrodeposition, directly between Zn metal and the substrate. This portion of alloying reaction is not captured by the capacity above 0 V v.s. $Zn^{2+}$/Zn in the plating branches of the galvanostatic voltage profiles. Also noticeable is the presence of a secondary plateau in the stripping voltage profile. The secondary plateau corresponds to the dealloying process, which occurs after the stripping of elemental zinc. These identified signatures of the interfacial reaction are in good agreement with the CV results (**Fig. 2**) and the XRD results (**Fig. 3**, **Fig. S10**); see also XPS in **Fig. S11**. Focused ion beam (FIB) and advanced aberration-corrected atomic resolution scanning transmission electron microscopy (STEM) were used to further characterize such alloying phenomena (**Fig. 5 A~C** and **Fig. S12~15**). Additionally, as shown in **Fig. S16~18**, we performed Zn plating/stripping efficiency measurement on a non-planar stainless-steel mesh to evaluate the possible influence from substrate roughness. The result shows that roughness on these scales have at most a negligible influence on Zn plating/stripping, in comparison with the interfacial chemistry.

**Discussion**

Taken altogether, our findings indicate that chemical interaction at the heterointerface is necessary to effectively prevent the detachment of metal deposits, in analogy to the first component to the Sabatier principle—that is, moderate chemical interaction is needed to promote adsorption of the molecules to the catalyst surface. We believe that the alloying-induced adhesion is comparable to the concept of "*diffusion welding*" in metallurgy (*36, 37*). Such diffusion welding is thermodynamically driven by alloying reactions—*i.e.*, the formation of metal-metal solid solution and/or intermetallics. Unlike a conventional thermal- or pressure-driven diffusion welding process, diffusion in this scenario is mainly driven by spontaneous chemical reaction. This is allowed due, in part, to the relatively low melting temperature of Zn ($T_m$=698 K). At room temperature, Zn has a moderate homologous temperature $T_H = \frac{T}{T_m} = 0.4$, suggesting that interdiffusion of Zn can occur on moderate time scales (*38*). Motivated by this analysis, we summarize the melting temperatures and the $T_H$ values of metals of contemporary interest as battery anodes (see **Table S1**). Our findings imply that the diffusion welding mechanism is likely to play an even more important role for the alkali metal electrodes, and possibly some role for Al electrodes.

One may naïvely conclude that a stronger alloying interaction would provide a greater driving force for the interdiffusion, and further suppress the detachment of metal deposits. We interrogated the reversibility of Zn electrodeposition on three metal substrates, Cu, Ag and Au, which all show some chemical interaction with Zn, but quite different electrochemical characteristics. The capacity contributed by the secondary plateau in the stripping process is used as a parameter for comparison across Cu, Ag and Au; 0.08, 0.23 and 0.49 mAh/cm$^2$ capacities (accounting for 10%, 25% and 70% of the total stripping capacity) and are attributed to dealloying reaction of Zn from Cu, Ag and Au, respectively. Interestingly, the amount of this secondary capacity is negatively correlated to the reversibility (*i.e.*, CE) and the cycle life achieved on the three substrates. This explicitly contradicts the conventional wisdom of a monotonic dependence of metal plating/stripping performance on the strength of the heterointerfacial chemical affinity. More insights can be obtained by comparing the galvanostatic plating/stripping efficiencies to the results obtained from the CV and XRD experiments discussed earlier. The significantly greater dealloying capacities of Ag and Au substrates are indicative of the bulk conversion, while the interaction between Zn and Cu is limited to the surface.

This last inference is supported by SEM characterization of the surface topology of the cycled substrates (**Fig. S19**)— Cu remains almost intact, while Ag and Au develop into porous structures. We note that this finding is consistent with the observation of a considerable amount of pulverized AuZn intermetallic phase stuck in the separator (**Fig. S7~9**). It means that a strong chemical interaction will result in bulk phase transition, which pulverizes the substrate, causes incomplete dissolution of the deposited metal (*e.g.*, see **Fig. 5A~C**; incomplete Zn dissolution only found on Ag but not on Cu after cycling ), and ultimately leads to battery failure. The extreme cases are demonstrated by Si, Sn, Ag and so forth when they are paired with Li metal (*39-41*). A large initial irreversibility and quick capacity fading is consistently observed. The optimal case in the Zn system is exemplified by Cu, where the substrate offers what appears to be a *just right*, moderate degree of interaction at the heterointerface with the metal deposits—neither too weak, which causes detachment of dead metal, nor too strong, which induces phase transitions penetrating the bulk. The pseudo-capacitive behavior in the underpotential regime (*i.e.*, >0 V v.s. Zn) of a negatively-polarized voltage sweep provides an easily accessible signature of substrates where such limited, interfacial chemical interaction exist and presage high levels of reversibility in metal anodes of batteries. These analyses could be used to interpret the observed variation of reversible plating/stripping behaviors over substrate chemistry (**Fig. 5D** and **Fig. S20**). We performed additional electrochemical measurements to evaluate the possible influences of surface oxides layer, desolvation structure and electrolyte decomposition on plating/stripping behaviors; the results suggest that they do not act as the determinant factor in scenarios studied in this work (**Fig. S21~23**).

In the original *Sabatier* principle for heterogeneous catalysis, formation enthalpy of the adsorbed species is used as a representative parameter for qualitatively evaluating the interaction strength. A more negative enthalpy means more heat is released by the reaction, and therefore indicative of stronger interactions. Here, an obvious, phenomenological relation can be seen—the dependence of plating/stripping reversibility on the capacity contributed by the secondary reaction (**Fig. 5E**); both the case of no secondary reaction and excessive secondary reaction are clearly unfavorable. We note that use of *secondary capacity* as a phenomenological parameter is also advantageous as it reflects the fundamental thermodynamic and kinetic parameters intrinsic to the chemistry.

Reorganizing the results using this scheme, we observe a similar volcano-like qualitative trend of electrochemical plating/stripping reversibility/stability over the range of formation enthalpy of the intermetallic species explored (**Fig. 5E**) (*12, 23, 42, 43*). Ta is a widely-known refractory, inert metal, which has been used to make crucibles to contain Zn for identifying the binary phase diagrams of Zn (*e.g.*, U-Zn, Mn-Zn, *etc.*). We therefore assumed that the interaction between Ta and Zn are thermodynamically unfavorable at room temperature, suggesting a positive formation enthalpy of Ta-Zn intermetallics. Judging from the electrochemical performance in **Fig. 4** and the visualization results in **Fig. S7~8**, the interaction between Ta and Zn is weaker than Cu-Zn, but stronger than SS-Zn and Al-Zn. As semi-quantitatively shown in the plot, SS (Fe, Cr), Al and Ta have positive formation enthalpies with Zn; Cu, Ag and Au have negative formation enthalpies with Zn, but the value for Au is approximately 4-5 times greater than Cu and Ag. The significantly greater heat dissipation generated by Au than by Cu and Ag, respectively, upon Zn deposition is experimentally confirmed by in *Operando* microcalorimetry (**Fig. S24**) (*44*). This analysis is in excellent agreement with the observed plating/stripping performance—Cu and Ag show much higher reversibility and stability than the metals on the left side (*e.g.*, SS, Ta, Al) and the metals on the right side (*e.g.*, Au). Interestingly, the electronegativity $\chi$ of these elements follows a sequence of Ta(1.5) < Ti(1.54) < Al(1.61) < Zn(1.65) < Cr(1.66) < Fe(1.83) < Cu(1.90) < Ag(1.93) < Au(2.54); this trend indicates that a greater $\Delta\chi = \chi_M - \chi_{Zn}$ may induce a stronger charge transfer effect between Zn and the substrate metal M and thus a stronger chemical bonding. We note further that in some other cases the alloying reaction between the metal deposits and the substrate could in theory be suppressed by surface passivation (*e.g.*, the presence of an oxide layer). This results in a scenario that falls onto the weak-interaction side. Viewed in a broader context, this framework can alternatively be used to comprehend other physical interactions between the metal deposit and the substrate. Earlier studies report that Zn exhibit extremely stable, reversible plating/stripping on graphene substrate (*12*). The adsorption energy of Zn onto graphene is -0.02 eV (*23*); very close to the optimal values identified in the present work.

The ease of tuning the reaction rate of the plating/stripping reaction by varying the externally imposed current offers an additional dimension for understanding heterointerfacial electrochemical phenomena in batteries. The ratio of the deposition rate and the alloy forming reaction rate is a natural dimensionless group to capture the influence of the current density. Based on the results in the preceding sections, one would expect that the negative effects of excessive alloying reactions observed on Ag and Au on electrode reversibility should be exacerbated at lower deposition rates but attenuated at higher rates. As shown in **Fig. S25**, the observations from experimental plating/stripping measurements are highly consistent with this expectation. At a low current density (*i.e.*, 2 mA/cm$^2$), Ag and Au exhibit more pronounced alloy formation behaviors as evidenced in the potential profile. This is accompanied by a clear deterioration in electrode reversibility as the current density decreases. At a high current density of 40 mA/cm$^2$, the traits of excessive alloying reactions are hardly observable, and the Coulombic efficiency values of the substrates are notably improved. An additional notable finding is that the plating/stripping remains highly stable on Cu at all current densities and areal capacities (see also **Fig. S26~27**). This suggests the interfacial nature of this process—underscoring the importance of our earlier observation that a moderately strong chemical interaction at the interface represents the optimal case for reversible plating/stripping of metals in battery anodes.

It is important to point out here that while the considerations based on energy (*i.e.*, formation enthalpy) appear to be fairly effective in qualitatively predicting the observed trends, the kinetic aspects—*e.g.,* the solid-state transport, the intrinsic chemical reaction rate, *etc*.—can play nontrivial roles. For example, the diffusivity of Zn in Ag is generally reported to be higher than that in Cu (*45-47*). This could originate from the larger atomic radius of Ag (Zn: 142 pm; Cu: 145 pm; Ag: 165 pm). The larger Ag lattice may allow additional diffusion mechanisms.

Notwithstanding the kinetical aspects, a reasonably good design rule can still be deduced from the data from an energy standpoint; Namely that a very negative formation enthalpy is consistent with a strong propensity of the system for undergoing bulk transition and invariably causes volume change which, after many plating/striping cycles, causes fatigue failure and pulverization of the substrate (*48*). This may in turn accelerate the transport kinetics. We further point out that the tendency for Zn to form solid-solution phases with Cu, Ag and Au could be an important kinetic step in the formation of intermetallic phases where the parent lattices of the substrates are initially preserved—in the solid-solution formation step—but later on transformed—in the intermetallic formation step. The absence of solid solutions in the phase diagram may result in sluggish kinetics that prevent such heterointerfacial alloying reactions from occurring at time scales relevant to general battery operation (*i.e.*, minutes to hours) at room temperature. The energy scale associated with solid solution formation—for example, between Zn and Cu—is on the order of negative 0.01 eV (*49, 50*); this means that alloying reactions of such a nature could also be sufficient to promote diffusion welding. A complete prediction may require detailed *ab initio* simulation to capture the kinetics step by step and the remaining aspects of the intermetallic phase formation, *e.g.*, volume change, elasticity, *etc*. The next step towards an atomistic, computational understanding of the system could be based on experimental measurements performed on single crystalline substrates.

We note further that, achieving highly stable and reversible Zn plating/stripping and leveraging it to advance the development of rechargeable batteries is of immediate interest to the growing community interested in low-cost, earth-abundant electrochemical couples capable of storing large amounts of electrical energy. A close to unity plating/stripping reversibility is required, for example, to create commercially relevant Zn battery systems (**Fig. S28**), because it minimizes the amount of Zn needed in the battery anode and lowers the amortized battery costs (*51*). As suggested by the plating/stripping CE results shown in **Fig. 4**, Cu stands out as a promising substrate for Zn anodes (see also **Fig. S29** for CE measured at 40 mA/cm$^2$: 99.94% over 10,000 cycles). Motivated by these observations, we assembled Zn full cell batteries by paring the Zn anode of interest with an imperfect, but state-of-the-art vanadium-based cathode (see **Fig. S30** for more details of the cathode design) (*52*). As a proof-of-concept demonstration, we deposit a known amount of Zn metal onto Cu and SS, respectively, and harvest them as two representative Zn anodes for full battery evaluation. We intentionally use a somewhat stringent N:P ratio of 3:1, which is approximately more than two orders of magnitude lower than conventional N:P values for Zn battery studies based on commercial Zn foil. Reversible metal plating/stripping on hetero-substrates is inevitable in these scenarios, *i.e.*, with low N:P ratios or even anode-free. The battery performance reported in **Fig. 6** shows that the Zn on Cu anode manifests stable cycling over 500 cycles with a 73% capacity retention (91% after 100 cycles), which clearly outperforms the Zn on SS anode (see also **Fig. S31**). It should be noted that the cathode material makes the dominant contribution to the observed capacity decay. More impressive is that the Zn on Cu electrode with a N:P ratio of 3:1 is no worse than commercial Zn foil with a N:P > 100:1. This observation means that *via* manipulating the chemical kinetics at the heterointerface, only <5% of the originally needed Zn is required here to achieve a same level of battery performance.

Due to the interfacial nature of the Sabatier-type principle, a thin coating of a rationally chosen substrate should be sufficient. Results in **Fig. S32~37** show that by coating a thin Cu skin layer (0.36 mg/cm$^2$) onto the SS substrate, the propensity of Zn for forming dead fragments observed on SS is completely suppressed. This means that essentially any substrate (probably chosen for other reasons, *e.g.*, lightweight) can be tailored using standard deposition methods to achieve favorable surface chemistry for the purpose of promoting highly reversible metal plating/stripping in battery anodes.

Finally, in light of the universality of the fundamental processes that underlie the proposed concept, we wanted to explore the relevance of the "Sabatier"-like principle to metal electrode reversibility in an entirely different context—*i.e.*, charge and discharge of sodium (Na) metal anodes. Na is, like Li, a bcc metal, highly reactive, and has a much lower reduction potential than Zn. Na is also mechanically fragile at room temperature and prone to orphaning (*52*), and is conventionally assumed to require formation of a stable SEI for reversible plating/stripping at planar substrates. We choose Au as a model substrate for Na deposition and in-depth analysis. Na-Au is attractive for fundamental and applications-related reasons. From a more fundamental perspective, DFT calculations report that its formation enthalpy per atom is -0.283 eV, which is comparable to the formation enthalpy of Zn-Au. This strong interaction reflects the large $\Delta\chi = 2.54 - 0.93 = 1.61$ between Na and Au. The large gap in electronegativity suggests a strong charge transfer effect between Na and Au when alloyed. From an applications perspective, Na is among the most promising candidates for electrical energy storage in the mass-manufactured, low-cost batteries needed to enable an "electrify everything" future.

In spite of the obvious differences between Zn and Na, we found that the Sabatier-like principle is relevant for designing substrates for achieving highly reversible Na metal deposition. Comparing Zn-Au and Na-Au, it is apparent and as expected that the XRD and electrochemical responses of Au to Na plating/stripping (**Fig. 7A~D**) show that Na-Au falls into the category of "too-strong" interaction, *i.e.*, the right side of the Sabatier-like relation (**Fig. 5**). The Sabatier-like framework offers a straightforward approach for securing highly reversible deposition of Na on Au: maintain the thickness of the Au layer below a certain value where the substrate bonding contribution is just right to prevent metal orphaning, but still weak-enough to enable complete recovery during the stripping process. In other words, the thickness of the Au substrate can serve as a highly effective knob for tuning the metal-substrate interaction, and therefore Na plating/stripping reversibility. The results reported in **Fig. 7D** clearly show that thinning down of the Au coatings to values in the range 10~100 nm drives the interaction into a more interfacial, stable regime, as evidenced by Na plating/stripping voltage profiles. These profiles are evidently similar to those reported earlier for Zn-Cu (**Fig. 4F**). **Fig. 7E** and **7F** show further that these changes also result in high Na plating/stripping reversibility, with an optimal CE value >99% achieved on the Au film with thickness of 50 nm. A more detailed discussion of these findings is provided in **Supplementary Note 1**.

In summary, we report a simple, yet powerful design principle for achieving reversible metal electro-plating/stripping behaviors at a heterointerface. Analogous to the Sabatier principle for heterogeneous chemical catalysis, we specifically find that a moderate interfacial chemical interaction strength between the metal and the substrate is favorable. This discovery is also in line with the recent progress of designing advanced Li-sulfur cathodes by achieving optimal interaction between the polysulfides and the host materials guided by a similar Sabatier-type relation. Under these conditions mechanical detachment of metal deposits (*i.e.*, dead metal formation) and excessive bulk phase transition, which pulverizes the substrate, can both be prevented. A rule-of-thumb criterion is that the formation enthalpy of the intermetallic species formed should be slightly negative; very roughly, this optimal value is estimated to be around -0.02 ~ -0.04 eV/formula unit. The study also demonstrates that advanced surface characterization tools (*e.g.*, FIB and aberration-corrected STEM) provide a robust approach for probing such critical electrochemical interfaces with high spatial and chemical resolution. The electroanalytical and structural characterization results together reveal the distinct natures of the chemical kinetics occurring at the heterointerfaces. We further show that this knowledge can be translated to guide the materials design of advanced battery anodes with much less need for excess metal in the anode, but which exhibit stable, long-term cycling behaviors.

**Materials and methods**

**Materials**

Zn foil (99.9%), $ZnSO_4 \cdot 7H_2O$ (99.95%), $V_2O_5$ (99.95%), NaCl (99%), $NaPF_6$ and glycol dimethyl were purchased from Sigma Aldrich. Ta foil (99.95%; 0.025 mm), Au foil (99.95%; 0.025 mm), Ag foil (99.95%; 0.025 mm), 304 stainless steel foil (0.025 mm), were bought from Alfa Aesar. Cu foil (99.8%; 0.025 mm) and Al foil (99.3%; 0.015 mm) were bought from MTI. The foils were used as received unless otherwise specified. Deionized water was obtained from Milli-Q water purification system. The resistivity of the deionized water is 18.2 MΩcm at room temperature. Zn electrolytes were prepared by dissolving $ZnSO_4 \cdot 7H_2O$ into the deionized water (2 M). Cu electrolytes were prepared by dissolving $CuSO_4 \cdot 5H_2O$ into the deionized water (1 M). Electrolyte for Na cells was prepared in the lab by dissolving 1.0 M sodium hexafluorophosphate ($NaPF_6$) in anhydrous, 99.5% diethylene glycol dimethyl ether, followed by further drying using molecular sieve. Plain carbon cloth 1071 HCB was purchased from Fuel Cell Store. Ketjen Black (KB) carbon was purchased from AkzoNobel. Stainless steel 304 discs with a diameter of 15.8mm were polished on-site by VibroMet vibratory polisher, using Final-POL polishing cloth and 0.3 um alumina slurry. Polished stainless steel discs were coated with a thin layer of gold (Au) by magnetron argon sputtering deposition system.

**Characterization of materials**

Field-emission scanning electron microscopy (FESEM) was carried out on Zeiss Gemini 500 Scanning Electron Microscope. Cyclic voltammetry was performed using a CH 600E electrochemical workstation. X-ray diffraction (XRD) was performed on a Bruker D8 General Area Detector Diffraction System with a Cu or a Co Kα X-ray source and a Bruker D8 powder diffractometer. Galvanostatic measurements were carried out using Neware battery testing system. Atomic force microscopy (AFM) was performed on Cypher ES (Asylum Research Inc.). Focused ion beam (FIB) was performed on the VELION system manufactured by Raith. A Zeiss Gemini scanning electron microscope (SEM) system was used to obtain the secondary electron images and the energy dispersive spectroscopy (EDS) results. Aberration-corrected high angle annular dark field – scanning transmission electron microscopy (HAADF-STEM) was carried out on FEI Themis Z G3. The materials characterization was performed in part at MIT.nano Characterization Facility.

X-ray Photoelectron Spectroscopy (XPS) data was collected at the Center for Functional Nanomaterials at Brookhaven National Laboratory, NY, in an ultrahigh vacuum chamber equipped with a SPECS Phoibos 100 MCD analyzer and non-monochromatized Al Kα (hν = 1486.6 eV) X-ray source. The experiments were carried out using an accelerating voltage of 12 kV and a current of 20 mA, with a base chamber pressure of $2 \times 10^{-9}$ torr. Cycling was terminated with a deposition step to a capacity of 0.1 mAh/cm$^2$. For all cycled samples, data were collected in the as-prepared state as well as after 45 minutes of Ar sputtering, performed at room temperature with a pressure of $2 \times 10^{-5}$ torr with an energy of 1.5 keV. Electrodes were pressed onto a conductive copper tape and mounted on the sample holder. Charge correction was performed by calibrating the Zn 2p binding energy to 1022 eV for Zn 2p$_{3/2}$ spectra for cycled samples and 284.8 eV for adventitious carbon1 for pristine foils. Data were analyzed using CASA XPS software. A Shirley background was subtracted prior to peak deconvolution. Fitting was performed using a mixed Lorentzian-Gaussian line shape.

The operando isothermal microcalorimetry (IMC) measurements were performed with a TA Instruments TAM IV microcalorimeter. Coin-type cells were placed in IMC ampules submerged in an oil bath where the temperature was rigorously maintained at 30 °C. All operando electrochemistry tests were controlled by a BioLogic VSP potentiostat. The cells monitored by the IMC were charged or discharged at a constant current of 0.8 mA/cm$^2$ for five cycles. A voltage limit of 1.0 V (vs Zn/Zn$^{2+}$) was used for metal stripping (charge, oxidation), while an areal capacity limit of 0.8 mAh/cm$^2$ was used to control the metal plating (discharge, reduction). Between each plating/stripping step, a one-hour open-circuit rest was introduced to ensure thermal equilibration.

**Electrochemical measurements**

**Fabrication of the cathode:** NaV$_3$O$_8$·1.5H$_2$O (NVO) was synthesized according to a prior study (*53*). One gram of V$_2$O$_5$ powder was mixed with 15 mL of 2 M NaCl solution. The mixture was stirred for 96 hours at room temperature. The product was washed and collected after freeze drying. The NVO material was mixed with ketjen black and polyvinylidene fluoride (weight ratio 80:10:10), and dispersed into carbon cloth following a procedure reported in our previous study (*52*). **Coin cell fabrication**: CR2032 coin cells were used. Electrodes are separated by glass fiber (Whatman) filter membranes. The thickness of a piece of free-standing glass fiber separator before battery assembling is on the order of hundreds of microns (200~400 μm). The effective thickness of the membranes is significantly reduced under pressure when used in coin cells. In each cell, ~100 uL electrolyte was added by pipette. **Cu deposition on stainless steel**: The electrodeposition of Cu on stainless steel was performed at a constant potential at - 0.5 V v.s. Cu$^{2+}$/Cu in coin cells. A piece of copper foil was used as the reference/counter electrode; a stainless steel was used as the substrate (working electrode). The areal capacity of deposited Cu is 0.3 mAh/cm$^2$, corresponding to a ~400 nm theoretical thickness. **Coulombic efficiency measurement:** see **Fig. S2** for details. The metal plating/stripping Coulombic efficiency (CE) = $\frac{stripping\ capacity}{plating\ capacity\ on\ the\ substrate} \times 100\%$, which quantifies the reversibility of the metal anode. For example, CE=100% means all the plated Zn on the substrate can be stripped; while CE=80% means that 80% of plated Zn can be stripped and 20% Zn is electrochemically inactive.

**ACKNOWLEDGEMENTS**

The authors express their gratitude to Dr. M. Pfeifer, Prof. X. Ren and Dr. R. Luo for valuable discussions. **Funding**: This work was supported as part of the Center for Mesoscale Transport Properties, an Energy Frontier Research Center supported by the U.S. Department of Energy, Office of Science, Basic Energy Sciences, under award #DE-SC0012673. This work made use of the Cornell Center for Materials Research Shared Facilities which are supported through the NSF MRSEC program (DMR-1719875). This work made use of the MIT.nano Characterization Facilities (Gemini SEM, Raith VELION FIB-SEM, Themis Z G3). **Author Contribution**: L.A.A. directed the project. J.Z. and L.A.A. conceived and designed this work. J.Z. Y.D., J.Y. and T.T. performed the electrodeposition, electrochemical measurements and structure characterizations. W.L., P.W., X.T., D.C.B., K.J.T., E.S.T. and A.C.M. performed the XPS and IMC experiments. All the authors analyzed and discussed the data. J.Z., L.A.A. and Y.D. wrote the manuscript with important inputs from all the authors. **Data Availability**: All data needed to evaluate the conclusions in the paper are present in the paper and/or the Supplementary Materials. **Competing Interest Statement**: The authors declare no other competing interests.

# List of supporting materials

Supplementary Table S1. Melting temperature and homologous temperature of metals of battery interest.

Supplementary Figure S1. Surface topology characterization of the metal substrates.

Supplementary Figure S2. Illustration of the metal plating/stripping measurement.

Supplementary Figure S3. Zn plating/stripping measurement on Cu with an upper cutoff voltage of 0.5 V v.s. $Zn^{2+}/Zn$.

Supplementary Figure S4. Zn plating/stripping measurement on Ti foil.

Supplementary Figure S5. Zn plating/stripping measurements at 4 $mAh/cm^2$.

Supplementary Figure S6. Zn plating/stripping measurements at 0.07 $mAh/cm^2$ on a SS substrate.

Supplementary Figure S7. Microstructural characterization of the separator surface facing metal heterointerfaces with different chemistries.

Supplementary Figure S8. Macroscopic photos of the separator surface facing metal heterointerfaces with different chemistries.

Supplementary Figure S9. Characterization of a separator facing the Au substrate after 100 plating/stripping cycles.

Supplementary Figure S10. XRD of substrates after one-time Zn deposition.

Supplementary Figure S11. XPS Zn spectra for cycled (a) aluminum, (b) iron, (c) tantalum, (d) copper, (e) silver, (f) gold substrates and (g) Zn foil reference.

Supplementary Figure S12. Characterization of surface structure of Ag after Zn plating/stripping.

Supplementary Figure S13. Characterization of surface structure of Cu after Zn plating/stripping.

Supplementary Figure S14. FIB-STEM characterization of surface structure formed on Ag after Zn plating/stripping.

Supplementary Figure S15. In-depth STEM characterization of surface structure of Ag after 20 times of Zn plating/stripping.

Supplementary Figure S16. Microstructure of the stainless steel mesh.

Supplementary Figure S17. Zn plating/stripping behaviors on stainless steel mesh.

Supplementary Figure S18. Optical microscopy characterization of the separator facing stainless steel mesh after one-time Zn deposition.

Supplementary Figure S19. SEM images showing electrode morphology of Au, Ag and Cu, respectively, after 100 Zn plating/stripping cycles at 8 mA/cm$^2$.

Supplementary Figure S20. The average value and the standard deviation of the metal plating/stripping Coulombic efficiency of the substrates.

Supplementary Figure. S21. Plating/stripping on acid-washed Cu and Al substrates.

Supplementary Figure S22. Evaluation of electrolyte decomposition.

Supplementary Figure S23. Open circuit voltage (OCV) of electrodes with 0.1 mAh/cm$^2$ Zn deposition.

Supplementary Figure S24. Operando isothermal microcalorimetry (IMC) of electrochemical Zn plating/stripping on different substrates

Supplementary Figure S25. Galvanostatic electrochemical deposition/dissolution behaviors of Zn on substrates of different chemistry at different current densities.

Supplementary Figure S26. Zn plating/stripping efficiency on different substrates at 2 mA/cm$^2$ and 2 mAh/cm$^2$.

Supplementary Figure S27. Zn plating/stripping Coulombic efficiency at higher areal capacities.

Supplementary Figure S28. Calculated capacity retention of a Zn metal anode.

Supplementary Figure S29. Plating/stripping of Zn on Cu at 40 mA/cm$^2$.

Supplementary Figure S30. SEM characterization of the sodium vanadate (NVO) cathode.

Supplementary Figure S31. Full battery cycling performance of using "Zn on Cu" anode and NVO cathode.

Supplementary Figure S32. Electrodeposition of Cu onto stainless steel substrate.

Supplementary Figure S33. SEM characterization of Cu-coated stainless steel substrate.

Supplementary Figure S34. Optical microscopy (OM) characterization of Cu-coated stainless steel substrate.

Supplementary Figure S35. XRD pattern of Cu-coated stainless steel substrate.

Supplementary Figure S36. Plating/stripping of Zn on Cu coated SS.

Supplementary Figure S37. Optical microscopy characterization of the separator facing Cu-coated stainless steel foil after one-time Zn deposition.

Supplementary Note S1. Evaluation of Na metal plating/stripping.

# Supporting Information

# Design Principles for Heterointerfacial Alloying Kinetics at Metallic Anodes in Rechargeable Batteries

Jingxu Zheng[1,2,†], Yue Deng[1,†], Wenzao Li[3,4], Jiefu Yin[5], Patrick J. West[4,6], Tian Tang[1], Xiao Tong[7], David C. Bock[4,8], Shuo Jin[5], Qing Zhao[5], Regina Garcia-Mendez[5], Kenneth J. Takeuchi[3,4,6,8], Esther S. Takeuchi[3,4,6,8], Amy C. Marschilok[3,4,6,8], Lynden A. Archer[1,5*]

1. Department of Materials Science and Engineering, Cornell University, Ithaca, NY 14853, USA.

2. Department of Physics, Massachusetts Institute of Technology, Cambridge, MA 02139, USA.

3. Department of Chemistry, State University of New York at Stony Brook, Stony Brook, NY 11794, USA.

4. Institute for Electrochemically Stored Energy, Stony Brook University, Stony Brook, NY 11794, USA.

5. Robert Frederick Smith School of Chemical and Biomolecular Engineering, Cornell University, Ithaca, NY 14853, USA.

6. Department of Materials Science and Chemical Engineering, State University of New York at Stony Brook, Stony Brook, NY 11794, USA.

7. Center for Functional Nanomaterials, Brookhaven National Laboratory, Upton, NY 11973, USA.

8. Interdisciplinary Science Department, Brookhaven National Laboratory, Upton, NY 11973, USA.

*Corresponding author: laa25@cornell.edu

[†] These authors contributed equally to this work.

| Metal | Melting temperature /K | Homologous temperature at RT |
|---|---|---|
| K | 336.5 | 0.89 |
| Na | 371 | 0.8 |
| Li | 453.5 | 0.66 |
| Zn | 692.5 | 0.43 |
| Mg | 923 | 0.32 |
| Al | 933 | 0.32 |
| Ca | 1115 | 0.27 |

**Supplementary Table S1. Melting temperature and homologous temperature at room temperature of metals contemporarily being pursued as battery anodes.** As a rough rule of thumb, a higher homologous temperature, qualitatively, suggests a higher probability of (inter)-diffusion to occur—for example—in a creep process.

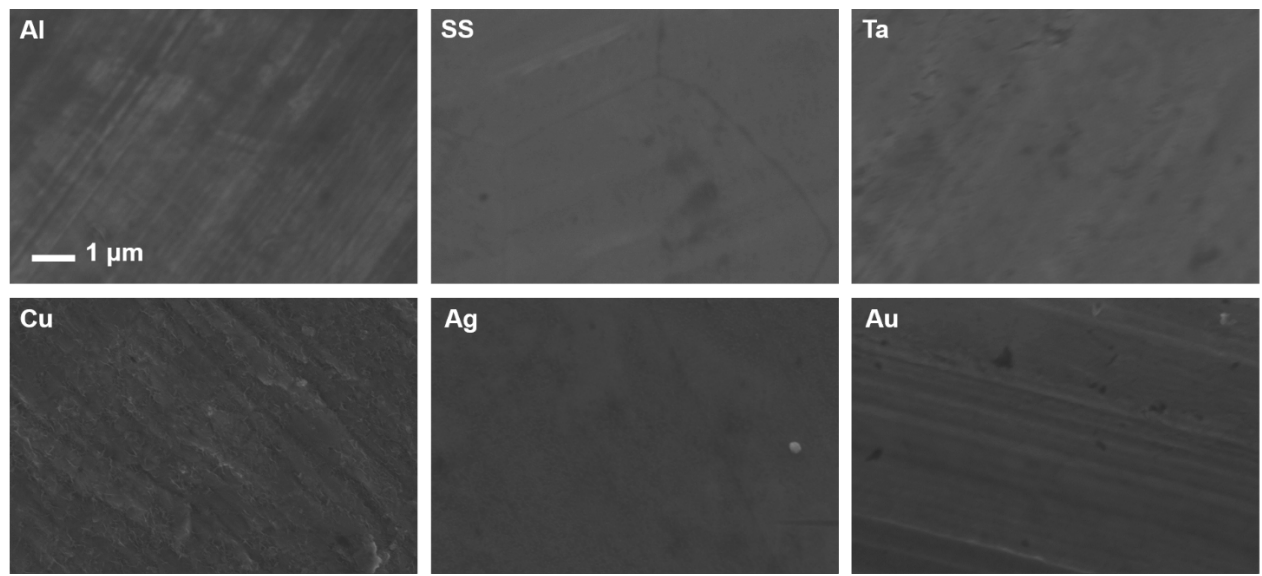

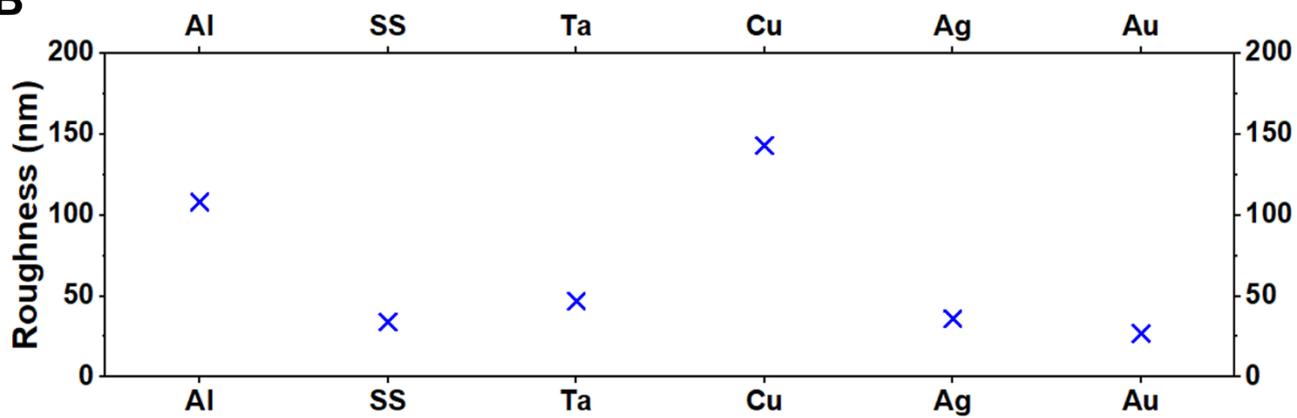

**Supplementary Figure S1. Surface topology characterization of the metal substrates.** (A) SEM images of the substrate surface (as-received) before electrochemical tests. These substrates consistently show a planar morphology. (B) Roughness of the substrates measured by AFM. The results show that these substrates all have comparable surface roughness values below 200 nm.

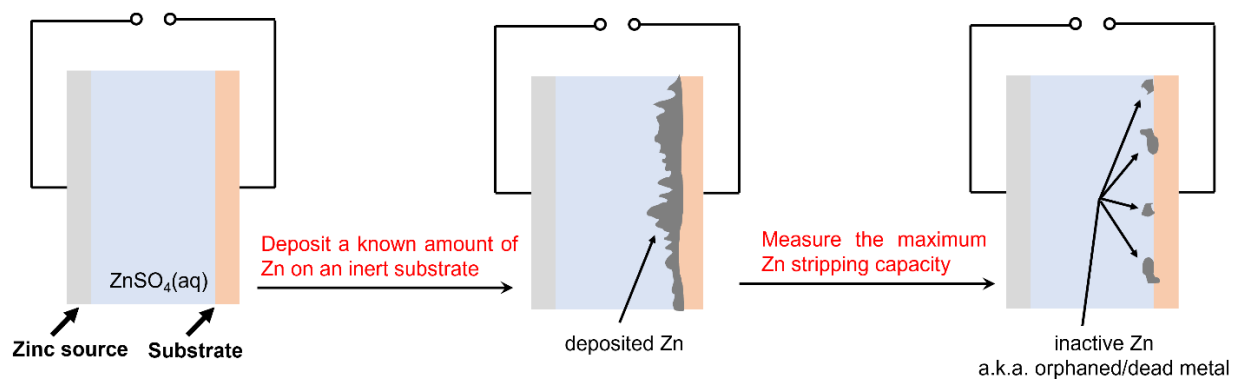

**Supplementary Figure S2. Illustration of the metal plating/stripping measurement.** A known amount of metal (*e.g.*, 1 mAh/cm$^2$ Zn) is first electro-plated onto the substrate under investigation. A reverse current is then applied to strip (*i.e.*, electrochemically dissolve/anodize) the deposits obtained in the previous step. The stripping process is cut off when the electrochemical potential of the substrate reaches a threshold (*e.g.*, 1.0 V v.s. Zn$^{2+}$/Zn). The Coulombic Efficiency (CE) is calculated according to CE=$\frac{Stripped\ capacity}{Deposited\ capacity}$ × 100%. It quantifies the reversibility of the metal plating/stripping process on the substrate. One major source of plating/stripping irreversibility is the formation of orphaned/dead metals that have no electrical connection to the current collector.

We point out that, any observed plating/stripping reversibility is determined by the amount of Zn that can be stripped at a certain rate; meaning that, it quantifies the amount of Zn that are "kinetically" active (54). The irreversible Zn deposits are "kinetically" inactive for a variety of reasons: (a) for orphaned/dead Zn, the characteristic time for electron transport is very large; for (b) for stranded Zn due to bulk conversion, the characteristic time for ion transport is very large (34). Put alternatively, in the thermodynamic limit (*i.e.*, t→∞), these kinetically inactive Zn still may be stripped, but this is clearly irrelevant in any practical batteries that charge and discharge within a certain period of time on the order of hours or shorter.

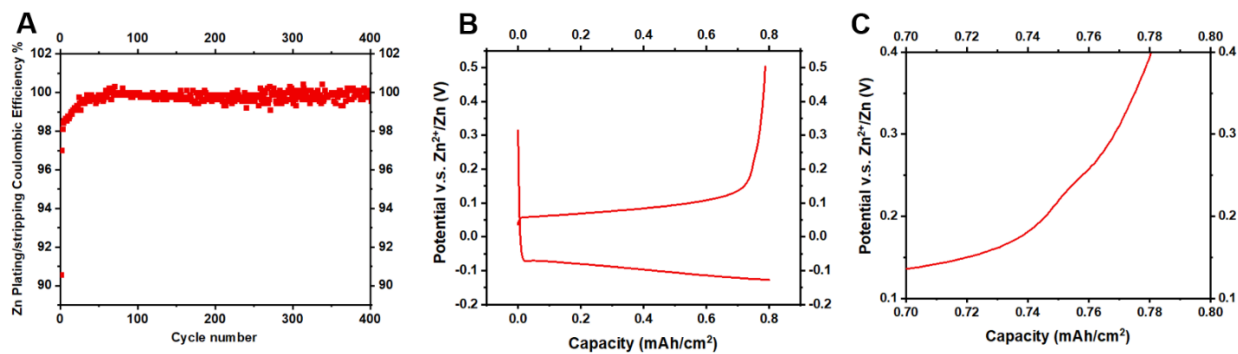

**Supplementary Figure S3. Zn plating/stripping measurement on Cu with an upper cutoff voltage of 0.5 V v.s. $Zn^{2+}$/Zn.** (A) Plating/stripping Coulombic efficiency values, (B) representative voltage profile and (C) a close-up view of the "kink" in the stripping branch, attributable to the dealloy process.

We note that the standard electrode potential of $Cu^{2+}$/Cu is 100 mV higher than the cutoff voltage (*i.e.*, 1.0 V v.s. $Zn^{2+}$/Zn) used in the measurements shown in **Fig. 4**. To rule out any possible influence from Cu oxidation/dissolution, we performed Zn plating/stripping on Cu foil using an upper cutoff voltage of 0.5 V v.s. $Zn^{2+}$/Zn, which is 600 mV below $\varphi^\circ(Cu^{2+}/Cu)$ and should avoid Cu dissolution. As evidenced in **Fig. S3**, the high Zn plating/stripping reversibility and the shape of the voltage profile observed under this condition are in line with the earlier results obtained with a +1.0 V cutoff voltage. The potentials at which Ag and Au anodization occurs are well above the cutoff voltage used in **Fig. 4**: Ag: +1.56 V, Au: +2.28 V v.s. $Zn^{2+}$/Zn.

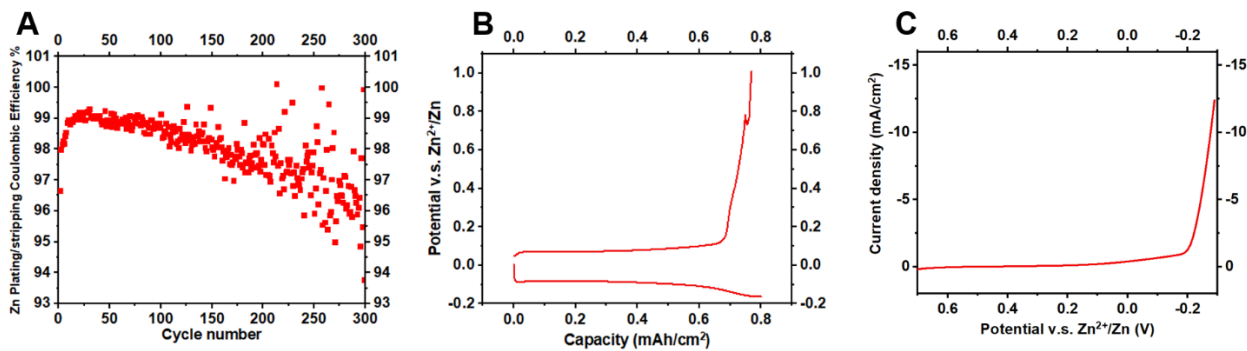

**Supplementary Figure S4. Zn plating/stripping measurement on Ti foil.** (A) Zn plating/stripping Coulombic efficiency and (B) voltage profile. (C) Linear sweep voltammetry on Ti foil (scan rate 0.1 V/s).

Consistent with literature reports (55), our results show that Ti foil falls into the category of (non-)/weakly-interacting substrate for Zn deposition. The Zn plating/stripping behavior on Ti foil features spiky voltage profiles and large scattering in CE values after cycling. This is similar to the observations made on other non-/weakly-interacting substrates, *e.g.*, Al/SS/Ta. No alloying reaction peaks are detectable in cyclic voltammetry scans.

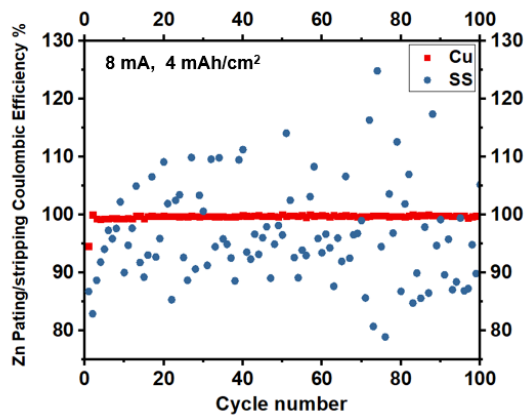

**Supplementary Figure S5. Zn plating/stripping measurements at 4 mAh/cm².** Cu and SS are chosen as representatives of interacting and non-interacting substrates, respectively, for evaluation at this high areal capacity.

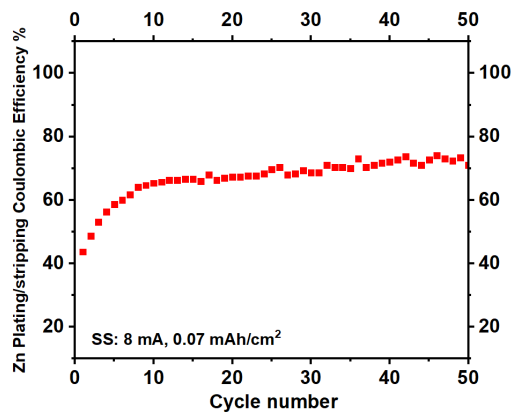

**Supplementary Figure S6. Zn plating/stripping measurements at 0.07 mAh/cm² on a SS substrate.** Current density 8 mA/cm². The smaller deviation of Zn plating/stripping Coulombic efficiency values suggests that the formation of dead metal is alleviated at this areal capacity that is one to two orders of magnitude smaller than practical values for batteries. Comparing this with measurements at higher capacities, It highlights the progressively critical importance of suppressing dead metal formation as the areal capacity approaches the regime practical for battery applications.

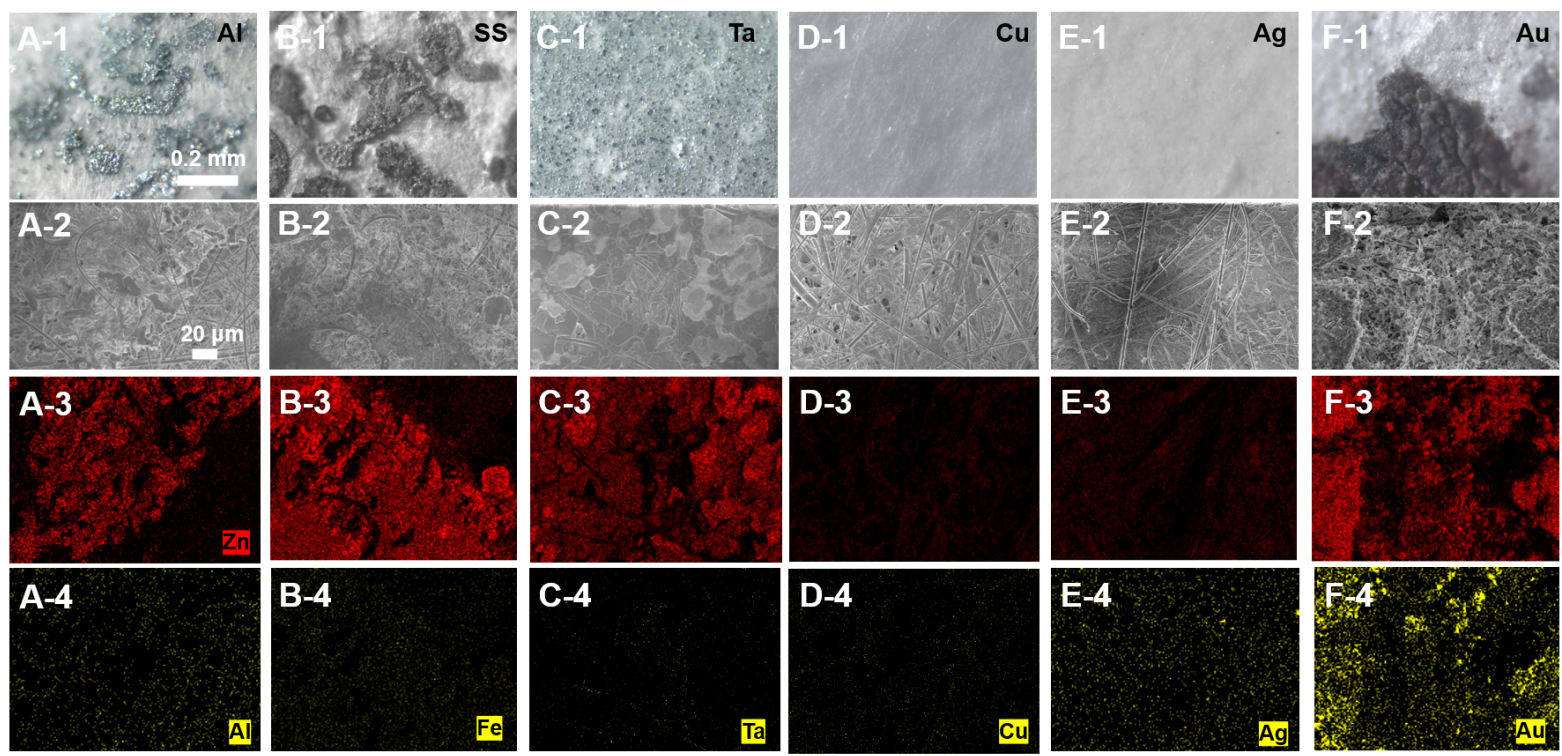

**Supplementary Figure S7. Microstructural characterization of the separator surface facing metal heterointerfaces with different chemistries.** Separators were harvested from cells using (A) Al, (B) SS, (C) Ta, (D) Cu, (E) Ag and (F) Au as substrates, respectively. Rows 1~4 summarize the optical microscopy images, the SEM images, and the corresponding EDS elemental mapping results of Zn and of the substrate metal, respectively. Large quantities of dead Zn fragments are observed to be stuck in the glass fiber separator facing stainless steel, Ta and Al. Pulverized intermetallic fragments (AuZn) are observed on the separator facing Au. The separators facing Cu and Ag, respectively, remain clean as shown in D-1 and E-1.

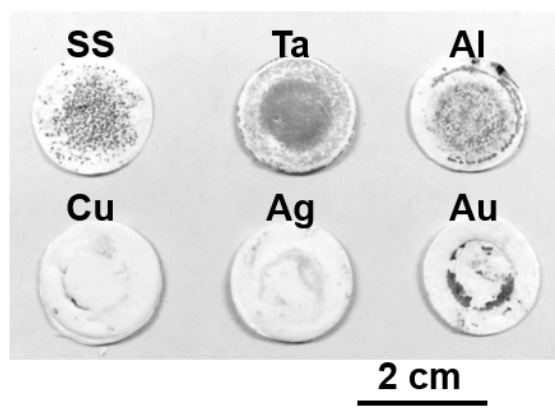

**Supplementary Figure S8.** Macroscopic photos of the separator surface facing metal heterointerfaces with different chemistries.

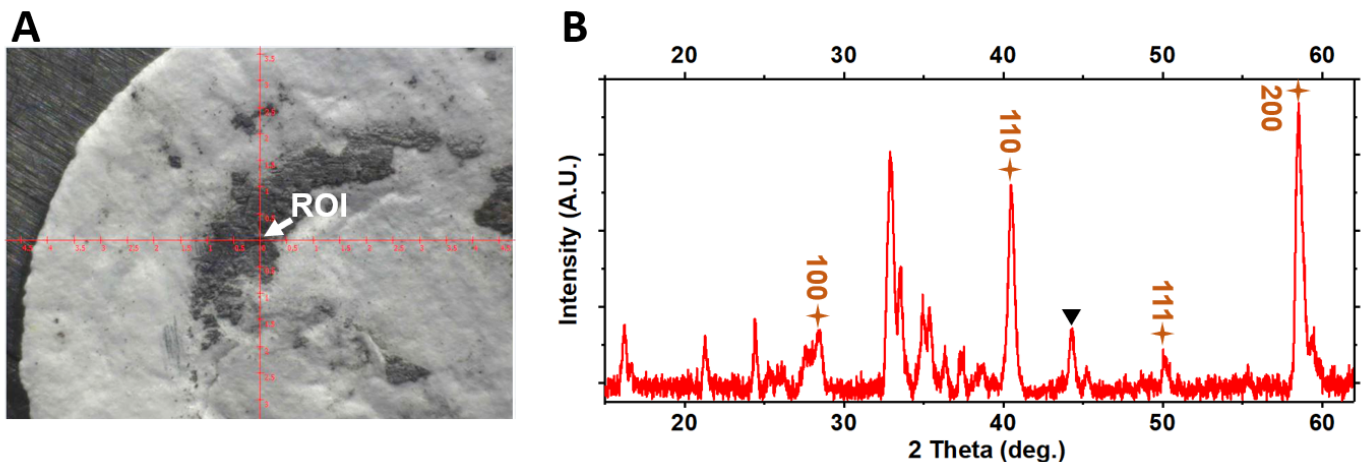

**Supplementary Figure S9. Characterization of a separator facing the Au substrate after 100 plating/stripping cycles.** (A) Optical microscopy image of the fragments stuck in the separator facing Au substrate. The arrow labelled by ROI indicates the location where the XRD pattern shown in (B) was collected. AuZn: brown cross ✚; Au: ▼ . The additional peaks are assigned to side products including $Zn_4SO_4(OH)_6·5H_2O$ (triclinic), whose presence suggests the high porosity of the fragments that traps the electrolyte upon battery dissembling (which can undergo hydrolysis when dried).

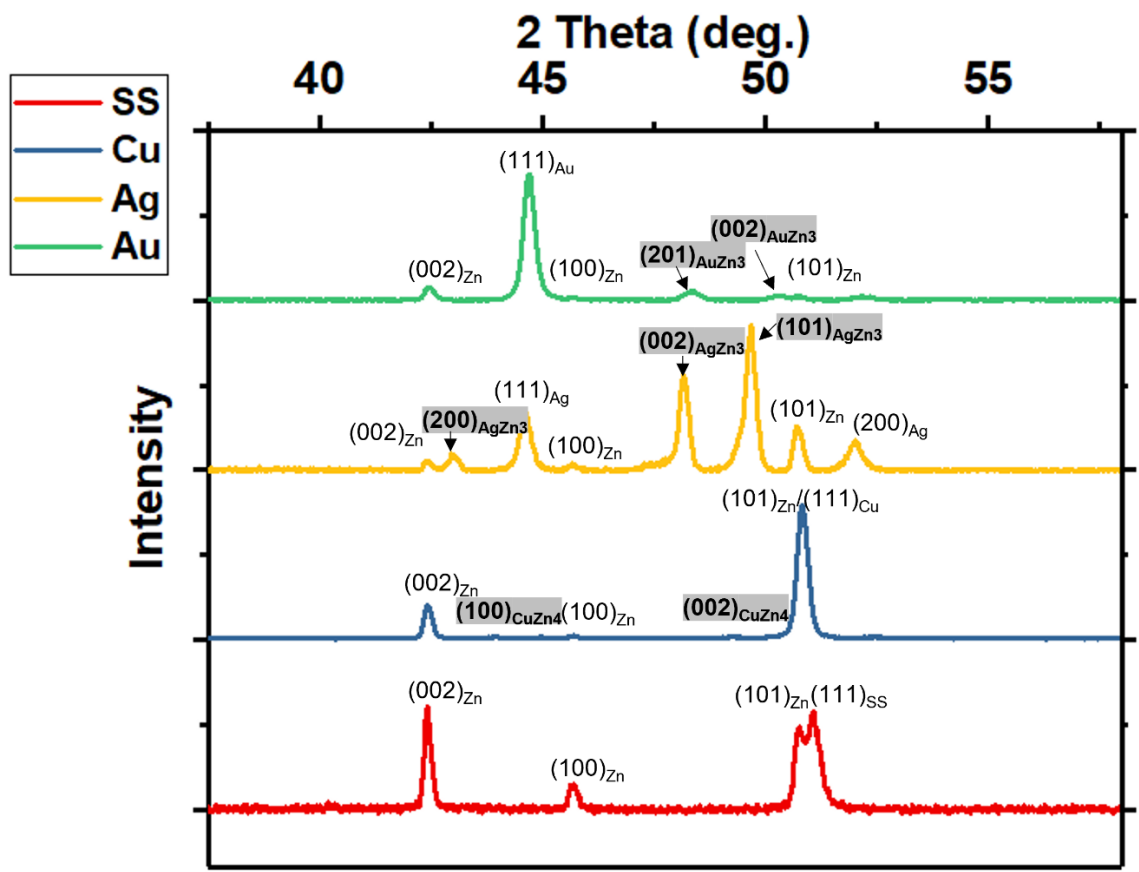

**Supplementary Figure S10. XRD of substrates after one-time Zn deposition.** Cobalt Kα X-ray source is used to collect the diffraction. In contrast to stainless steel, XRD peaks of alloys formed by Zn and substrates are observed on Cu, Ag and Au. The intensity of alloy peaks on Cu is relatively low because of the interfacial nature of the interphase. This observation is consistent with our conclusion that Ag and Au undergo a bulk conversion while the alloying on Cu is mainly interfacial. An O-Ring was used to separate the two electrodes to prevent the peel-off of deposits from the substrates as is seen when glass fiber is used as the separator. Condition: 8 mA, 0.8 mAh/cm$^2$.

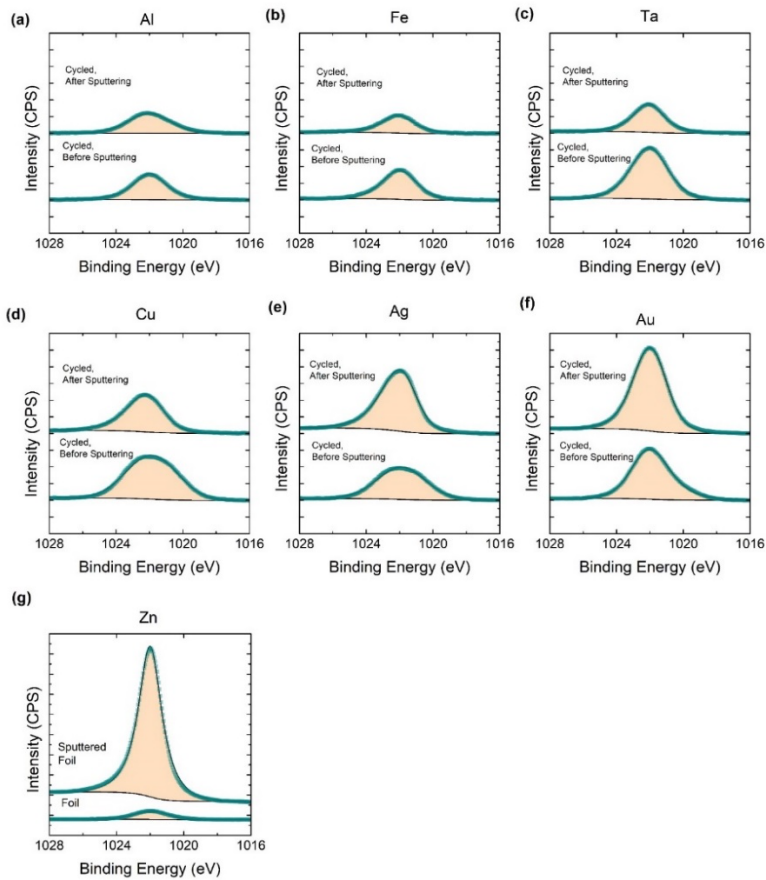

**Supplementary Figure S11. XPS Zn spectra for cycled (a) aluminum, (b) iron, (c) tantalum, (d) copper, (e) silver, (f) gold substrates and (g) Zn foil reference.** For each substrate, measurements before and after $Ar^+$ sputtering. For each substrate, measurements before and after $Ar^+$ sputtering. The results show that Zn mainly remains at an oxidation state of 0 (*i.e.*, $Zn^0$) on all these substrates with or without alloying reaction. For XPS analysis of the Cu, Ag, and Au foils, that showed evidence of alloying/dealloying during plating/stripping cycles, it is noted that the XPS technique cannot be used to reliably distinguish pure metals from metal alloys because of only low (0 - 0.5 eV) energy shifts upon alloying. For example, previous experimental results have shown that alloys of Zn and Cu cannot be resolved from their corresponding pure metals by either binding energies or Auger signal (56). Similarly, Ag and Au alloys also have XPS binding energies that are nearly equivalent to the corresponding pure materials, making discrimination between them problematic (57-59).

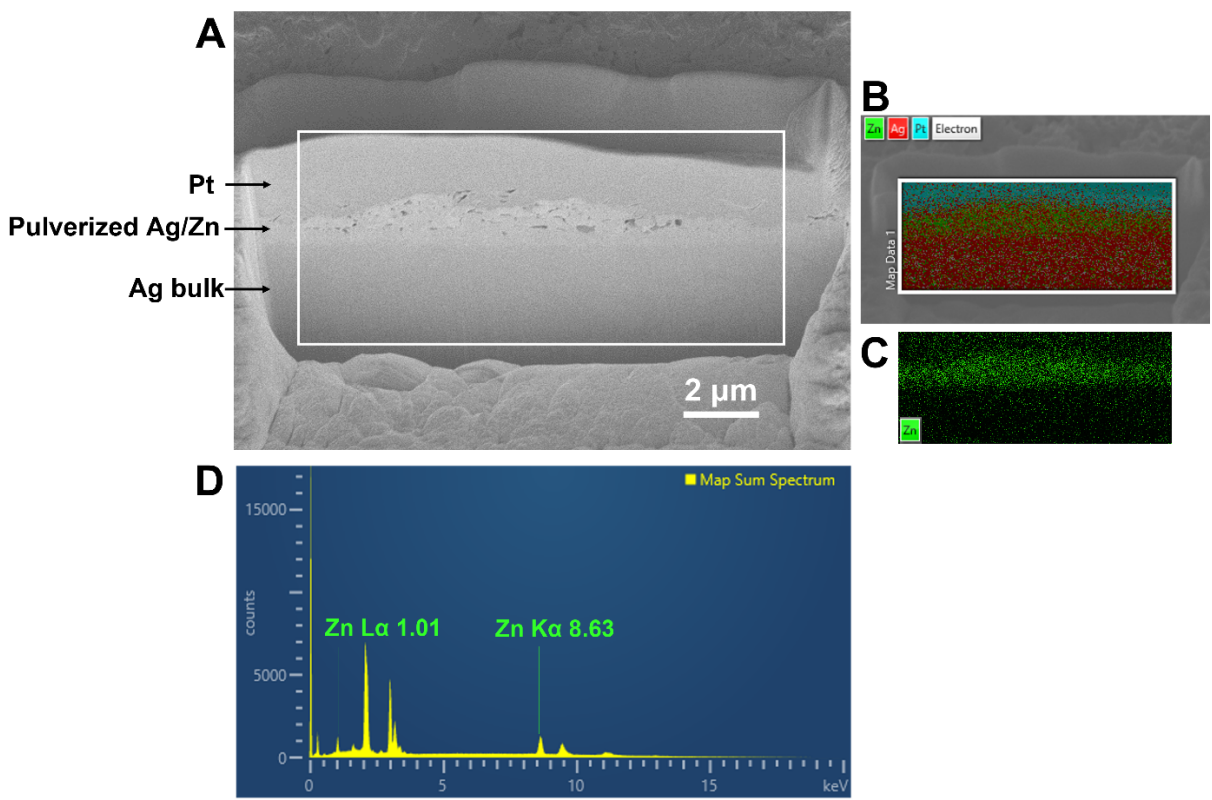

**Supplementary Figure S12. Characterization of surface structure of Ag after Zn plating/stripping.** Condition: 8 mA/cm$^2$, 0.2 mAh/cm$^2$; 100 cycles. (A) cross-section SEM image of the interface. (B)(C) EDS mapping results of the interface. (D) EDS profile of the interface.

By means of focused ion beam (FIB) cross-section cutting on Ag and Cu substrates after Zn plating/stripping (100 cycles) we interrogate the interphases formed in detail. Au substrate pulverizes into highly porous structures that do not allow such process due to unacceptably large roughness and porosity (**Fig. S19**). As can be clearly seen in **Fig. S12~13**, a Zn-enriched phase is found on Ag of a thickness around 1 micron after stripping to a cutoff of +1 V v.s. Zn$^{2+}$/Zn, suggestive of incomplete stripping, whereas no such Zn-enriched phase is observable on Cu substrate after plating/stripping, suggestive of nearly complete stripping of Zn.

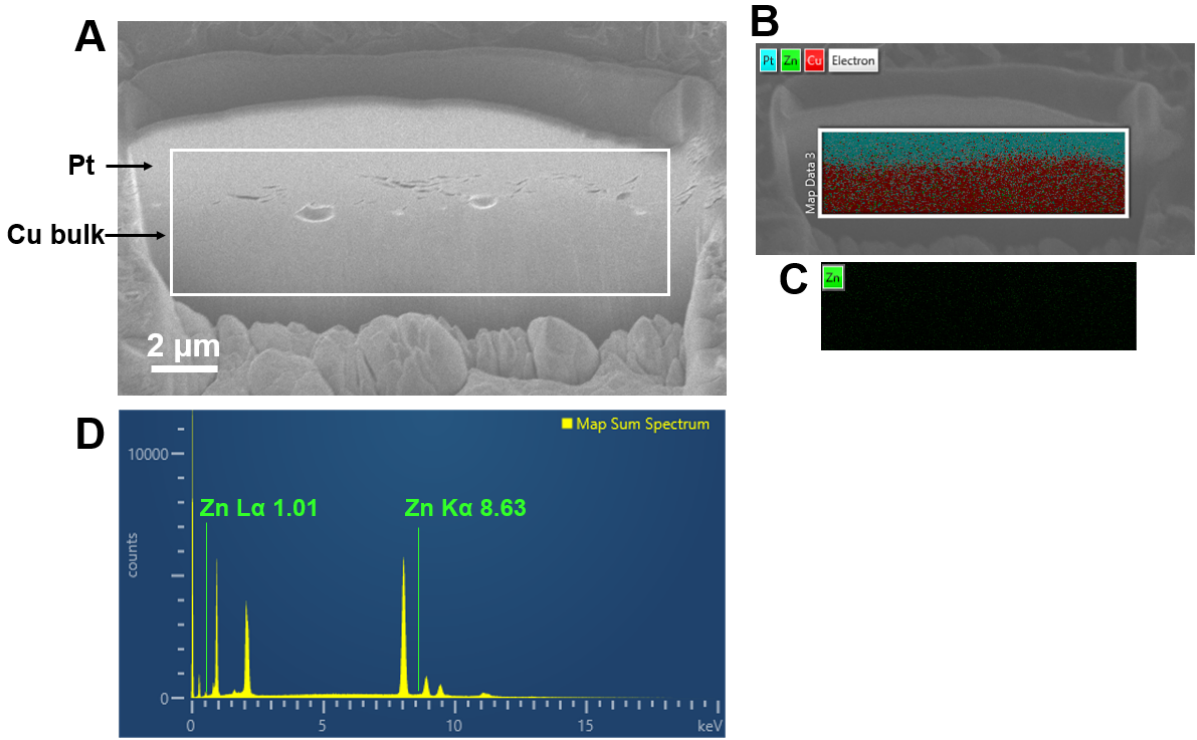

**Supplementary Figure S13. Characterization of surface structure of Cu after Zn plating/stripping.** Condition: 8 mA/cm$^2$, 0.2 mAh/cm$^2$; 100 cycles. (A) cross-section SEM image of the interface. (B)(C) EDS mapping results of the interface. (D) EDS profile of the interface.

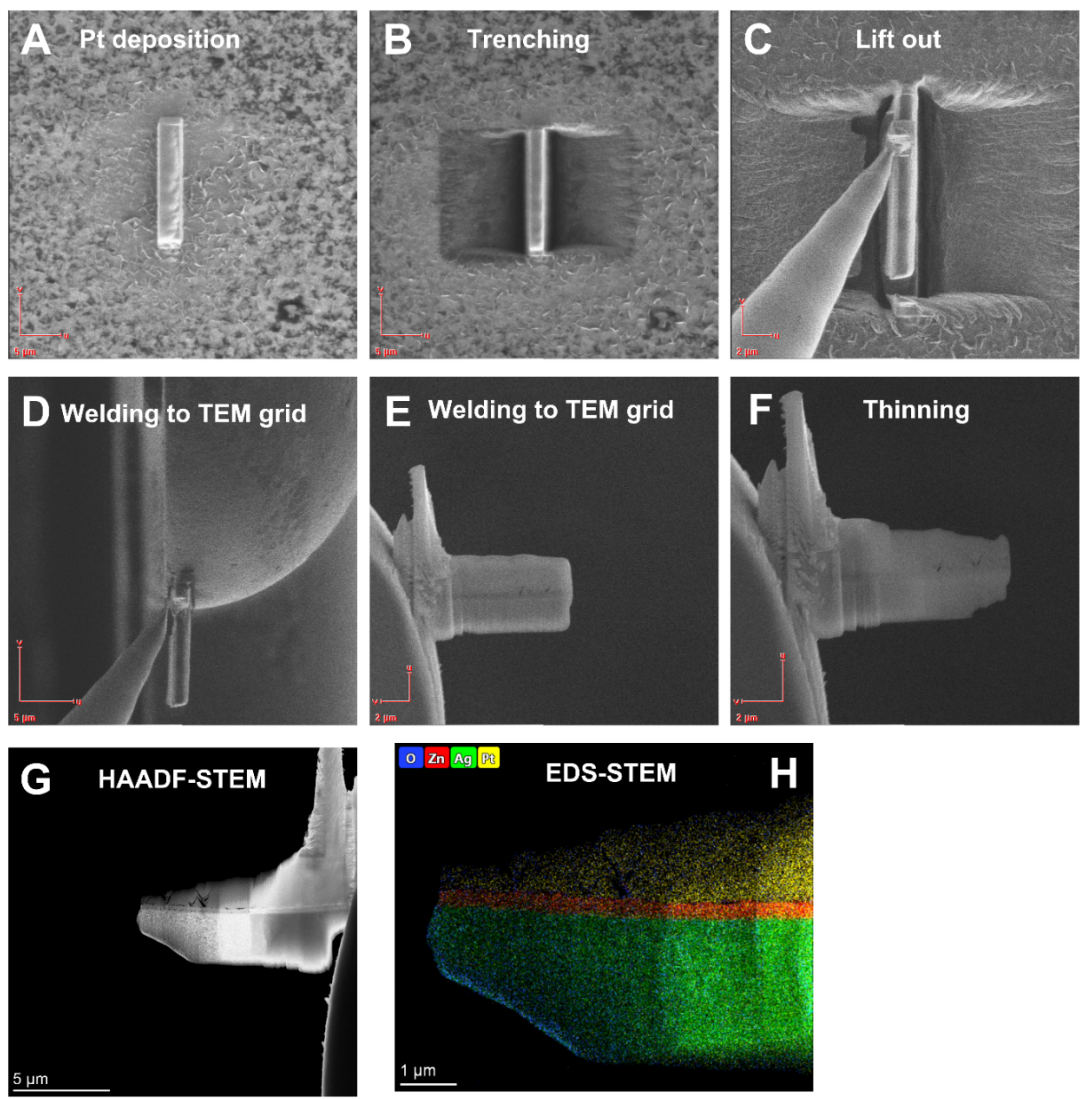

**Supplementary Figure S14. FIB-STEM characterization of surface structure formed on Ag after Zn plating/stripping.** (A)~(F) Preparation of TEM sample lifted out from a cycled Ag substrate. (G) HAADF-STEM and (H) EDS-STEM characterization of the lamella. Condition: 20 cycles; 8 mA, 0.2 mAh/cm$^2$; stripped state.

In light of the observations in **Fig. S12~13**, we performed TEM lamella lift-out process using FIB on the cycled Ag substrate (20 plating/stripping cycles, to make sure the surface alloy layer is not too thick for the process) to probe the precise crystallography and composition of the alloy phase associated with incomplete stripping.

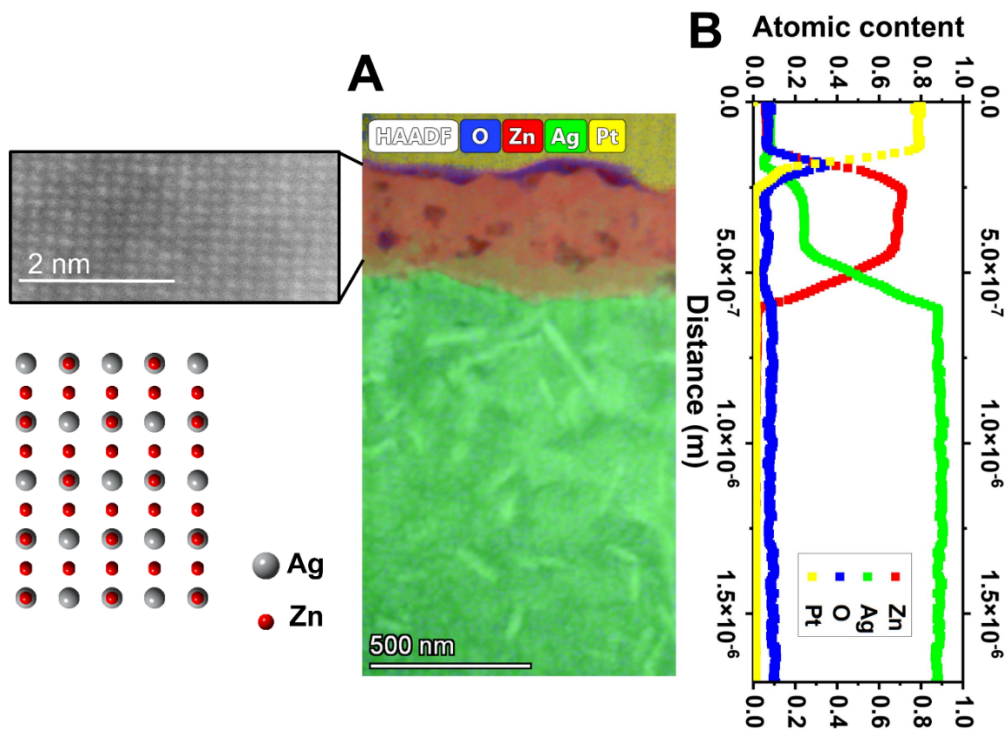

**Supplementary Figure S15. In-depth STEM characterization of surface structure of Ag after 20 times of Zn plating/stripping**. Condition: 8 mA/cm$^2$, 0.2 mAh/cm$^2$. (A) Overlaid HAADF/STEM image and EDS mapping results. The atomic-resolution HAADF-STEM image of the AgZn$_3$ alloy phase is shown to the right. (B) Atomic content values of relevant elements across the interface quantified according to the EDS spectra.

**Fig. S15** reports the results from aberration-corrected HAADF/STEM-EDS characterizations. An alloy phase enriched by both Ag and Zn is observed on the surface. Semi-quantitative EDS results show that the atomic ratio between Zn and Ag is 3:1, suggesting the formation of AgZn$_3$ phase. Impressively, this is in high consistency with our XRD diffraction results suggesting that the AgZn$_3$ is formed. As a step further, the atomic-resolution HAADF-STEM of alloy phase shown on the left side also confirms the atomic arrangements are consistent with the AgZn$_3$ phase. Together, these characterization results about the surface chemistry and crystallography with sub-angstrom spatial resolution using advanced electron/ ion microscopies provide strong support for our proposed concept.

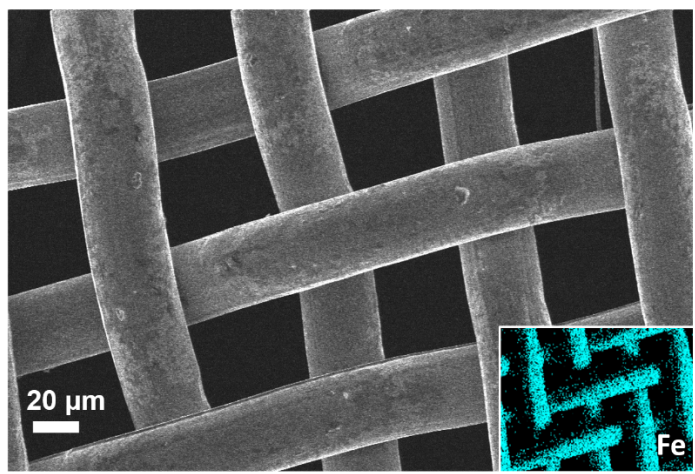

**Supplementary Figure S16. Microstructure of the stainless-steel mesh.** The inset shows the corresponding EDS mapping result.

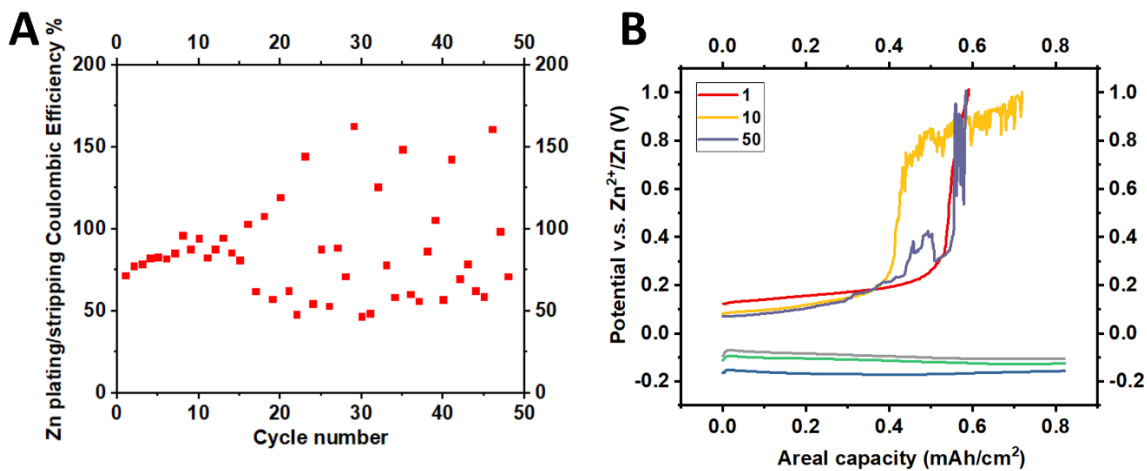

**Supplementary Figure S17. Zn plating/stripping behaviors on stainless steel mesh.** Current density = 8 mA/cm². (A) Coulombic efficiency and (B) representative potential profiles. The erratic CE values and potential evolution suggest that the Zn deposits on stainless steel mesh forms dead fragment easily, similar to stainless steel foil.

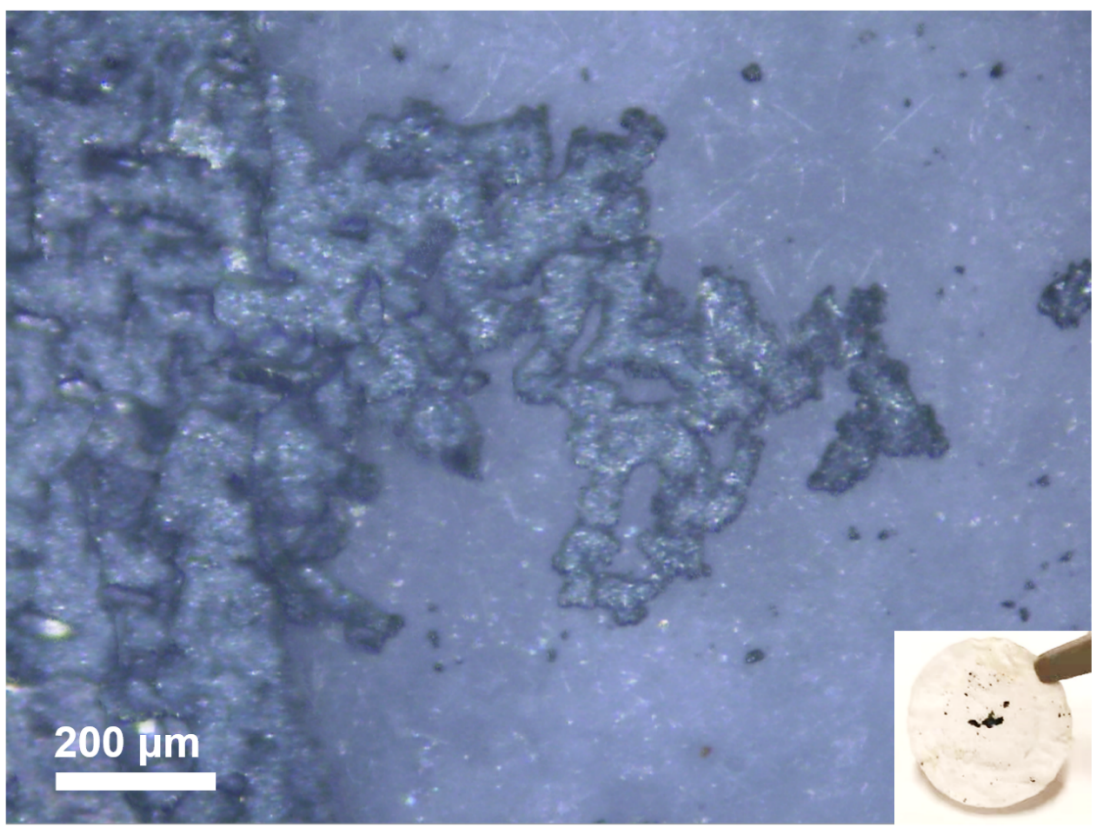

**Supplementary Figure S18. Optical microscopy characterization of the separator facing stainless steel mesh after one-time Zn deposition.** The deposition is performed at 8 mA/cm², 0.8 mAh/cm². Zn deposits on non-planar stainless steel detaches from the substrate easily, and are stuck in the separator. This observation is consistent with the electrochemical measurements in **Fig. S17**.

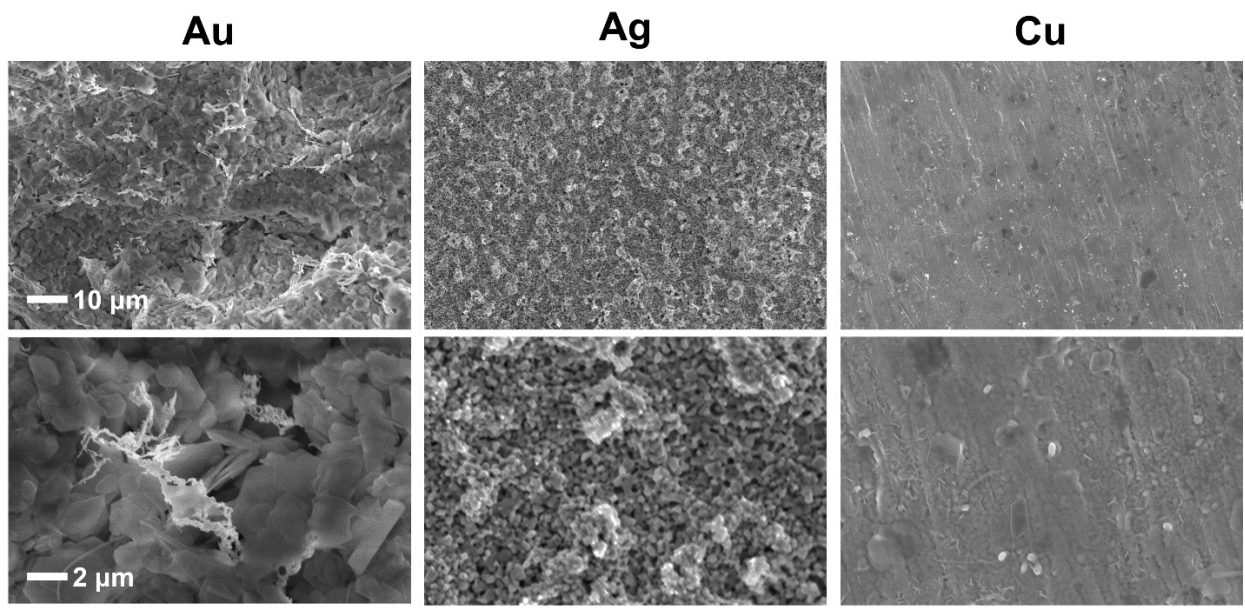

**Supplementary Figure S19. SEM images showing electrode morphology of Au, Ag and Cu foils, respectively, after 100 Zn plating/stripping cycles at 8 mA/cm$^2$.** Au and Ag substrates exhibit a highly porous morphology, whereas the Cu remains planar and compact, similar to the intact Cu electrode.

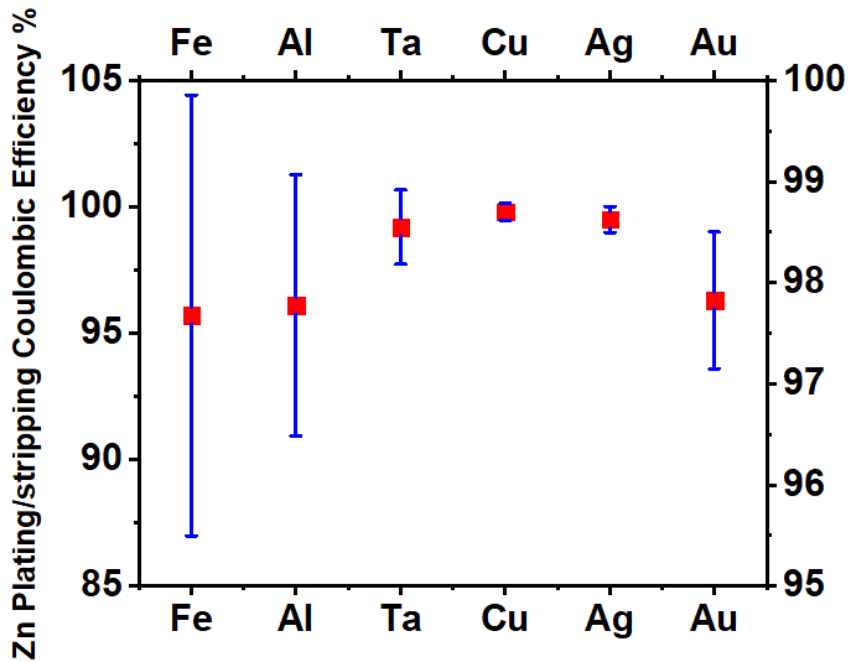

**Supplementary Figure S20. The average value and the standard deviation of the metal plating/stripping Coulombic efficiency of the substrates.**

While the CE difference between Cu and Ag seems to be relatively small when compared with other substrates, it can develop into significant differences over prolonged cycling. As a defining characteristic of rechargeable battery, it is a closed system and any instability/capacity fading is therefore accumulative over cycling.

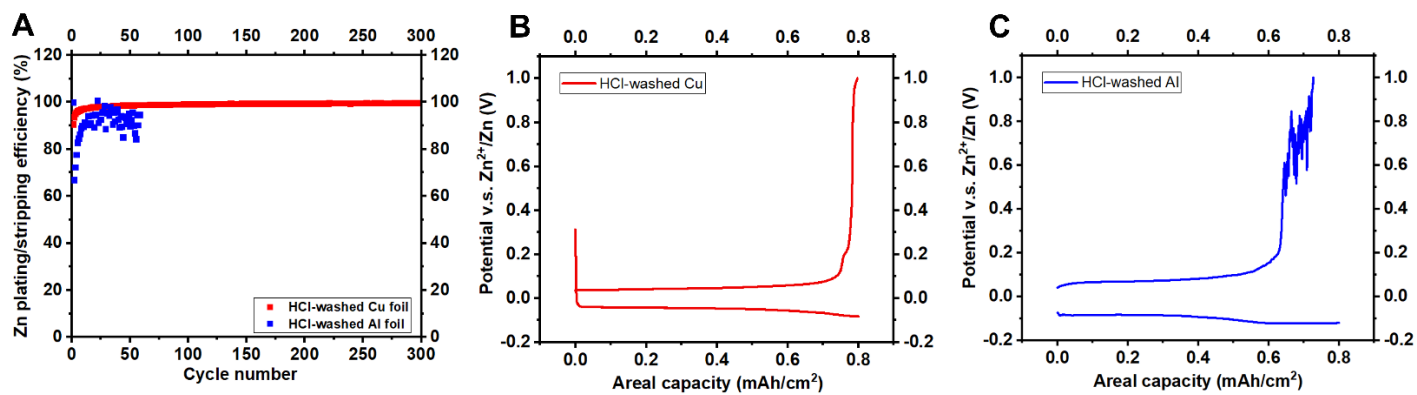

**Supplementary Figure. S21. Plating/stripping on acid-washed Cu and Al substrates.** (A) cycling performance; voltage profiles of Zn plating/stripping on HCl washed (B) Cu and (C) Al substrate.

To experimentally evaluate to what extent the surface oxidation could influence the plating/stripping behavior, we measured the Zn plating/stripping on substrates washed with dilute HCl, which effectively removes the surface oxides according to prior literature (60). For Al, the HCl reacts with the substrate and forms $H_2$ gas, confirming the exposure of metallic Al. The results show that such a surface chemical cleaning imposes very limited influences on the reversible plating/stripping behaviors. Our findings therefore suggest that the oxide does not play a dominant role for the substrates under the conditions used in the present paper.

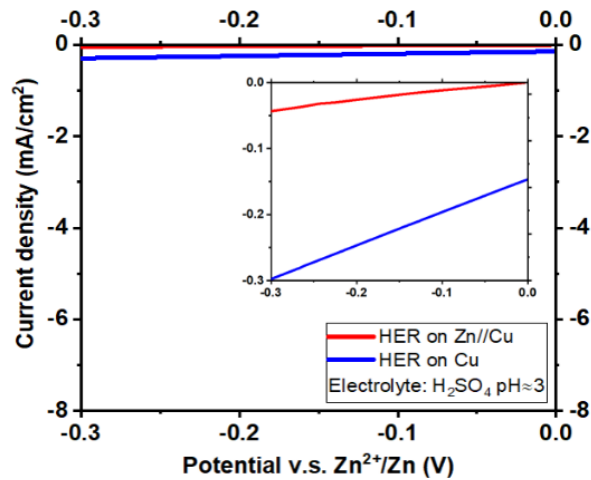

**Supplementary Figure S22. Evaluation of electrolyte decomposition.** Linear sweep voltammetry on bare Cu(blue) and Zn-coated Cu(red). On both substrates, the observed currents within the electrochemical window of interest are negligible, *i.e.*, <0.3 mA/cm$^2$. This means that the hydrogen evolution reaction (HER) on these surfaces exhibits sluggish kinetics, and significantly higher overpotentials are needed to drive HER with current densities that become nontrivial. After the deposition of the initial Zn layer on the substrate, the electrodes (regardless of underlying substrate material) are coated by Zn; the electrolyte decomposition occurs at the interface between electron conduction and ion conduction, which is the Zn// electrolyte interface. We note that the Zn-coated electrode shows even more suppressed water decomposition, due to the poor HER catalytic activity of Zn (fully filled 3d and 4s orbitals) (61). Particularly, after Zn deposition on electrode surface, the substrate underneath is no longer in contact with the electrolyte. In the stripping process, the substrate may be exposed to the electrolyte directly in the final stages of stripping. We would also point out that in the stripping process (*i.e.*, anodization), only reactions that are of an anodic/oxidative nature are allowed electrochemically. As such, HER is not allowed on the anode, but the corresponding side reaction is instead oxygen evolution reaction (OER). It is known that the onset of OER in aqueous electrolyte is at least > 2.0 V versus Zn$^{2+}$/Zn. This is much higher than the potential range that the electrode is cycled within. We therefore conclude that side reactions from water splitting should not play a prominent role in the current context.

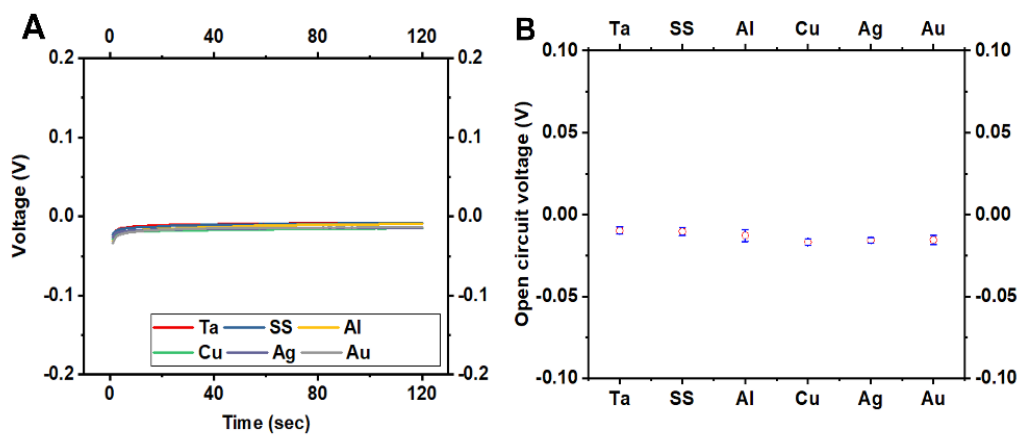

**Supplementary Figure S23. Open circuit voltage (OCV) of electrodes with 0.1 mAh/cm² Zn deposition.**

We measured the OCV of the cell after 0.1 mAh/cm² Zn deposition to evaluate any possible influence from the electric double layer and ion desolvation processes. In the present case, after the deposition of the initial Zn layers on the substrate, the electrode-electrolyte interface is in essence a conventional Zn metal || electrolyte interface, and the hetero-substrate (*e.g.*, Cu, Ag, Au) is not in direct spatial contact with the electrolyte. Experimentally, it is possible to probe this contact interface by measuring the OCV after depositing a small amount of Zn on the electrode surface. As shown in **Fig. S23**, the OCVs for the substrate with 0.1 mAh/cm² Zn deposition are very similar, suggesting negligible differences in the contact interface. We therefore conclude that the influence of hetero-interfacial interactions at the bottom of the metal deposits is negligible in terms of altering the electric double layer and the ion (de)/solvation process.

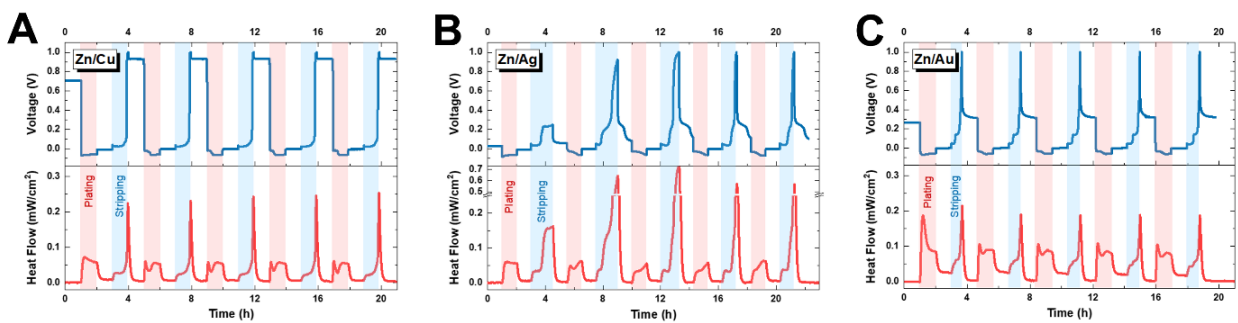

**Supplementary Figure S24.** *Operando* **isothermal microcalorimetry (IMC) of electrochemical Zn plating/stripping on different substrates**: (A) copper, (B) silver, and (C) gold. The heat dissipation observed on Au during Zn electrodeposition is approximately two times those observed on Cu and Ag. All plating/stripping processes were performed at a current density of 0.8 mA/cm$^2$. Between each plating (highlighted in red in each sub-plot) and stripping (highlighted in blue) step, a one-hour rest step was introduced.

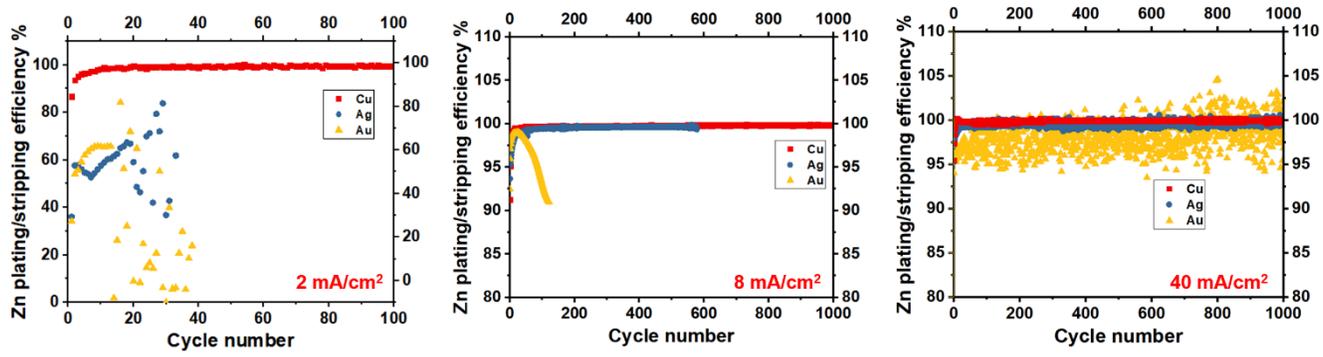
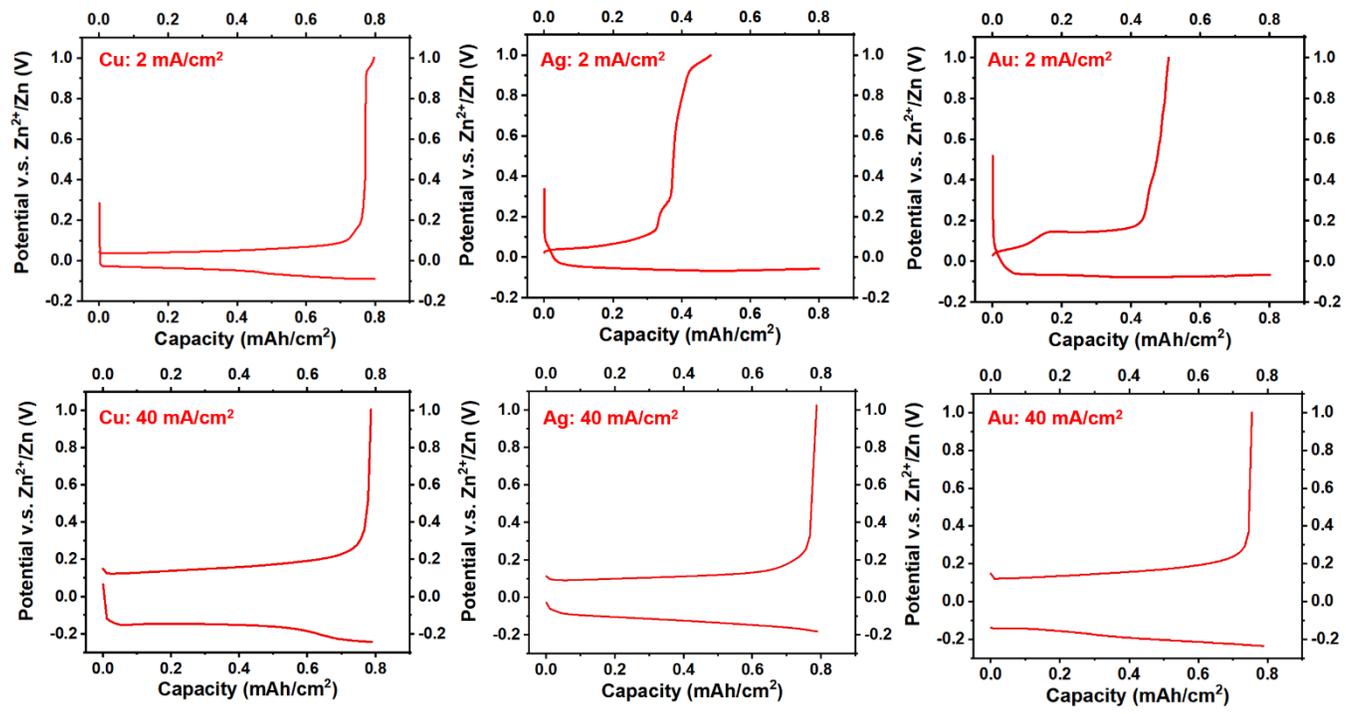

**Supplementary Figure S25.** Galvanostatic electrochemical plating/stripping behaviors of Zn on substrates of different chemistries at different current densities.

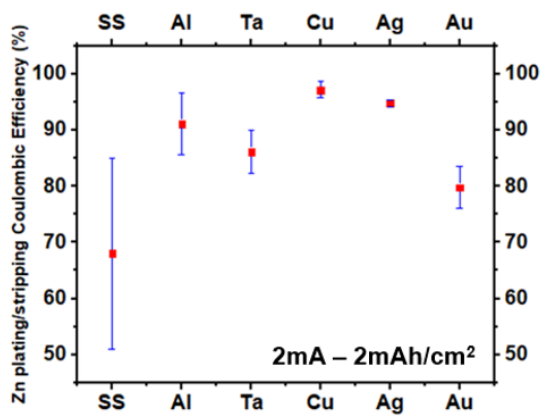

**Supplementary Figure S26.** Zn plating/stripping efficiency on different substrates at 2 mA/cm$^2$ and 2 mAh/cm$^2$.

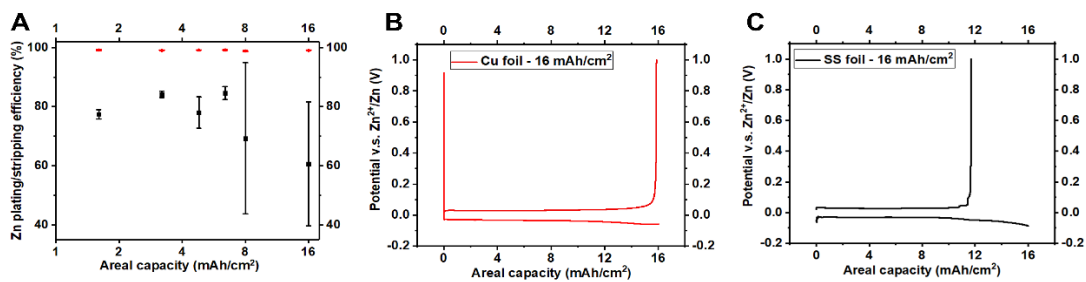

**Supplementary Figure S27. Zn plating/stripping Coulombic efficiency at higher areal capacities.** (A) Relation between Zn Coulombic efficiency and areal capacity. Voltage profiles of Zn plating/stripping on (B) Cu and (C) SS foil at an areal capacity of 16 mAh/cm$^2$. The substrate-deposit interface persistently plays a critical role in the reversible plating/stripping at elevated areal capacities. The reason behind can be seen by examining the stripping process—a non-interacting interface destabilizes the stripping process via detachment of the deposits at the root, *i.e.*, the substrate-deposit interface, regardless of the size/ thickness of the deposits.

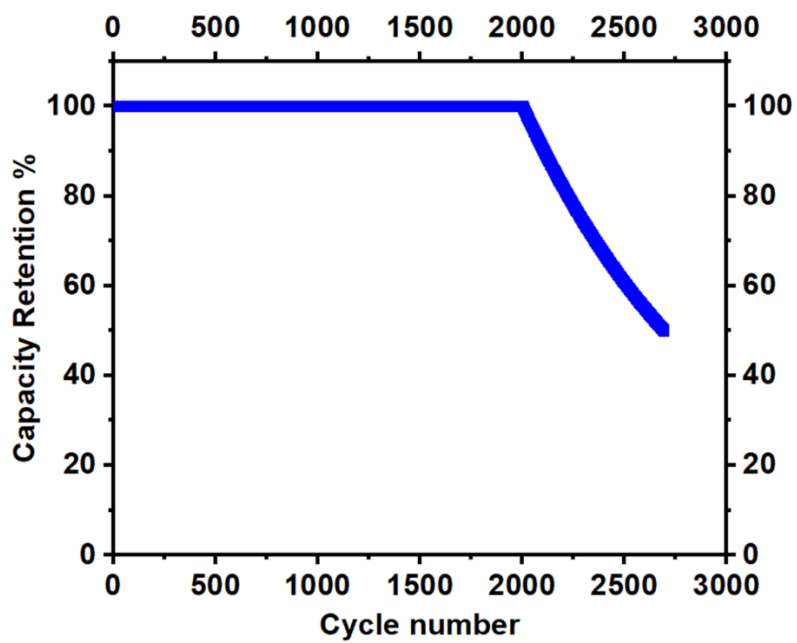

**Supplementary Figure S28. Calculated capacity retention of a Zn metal anode.** CE=99.9% and N: P = 3 : 1.

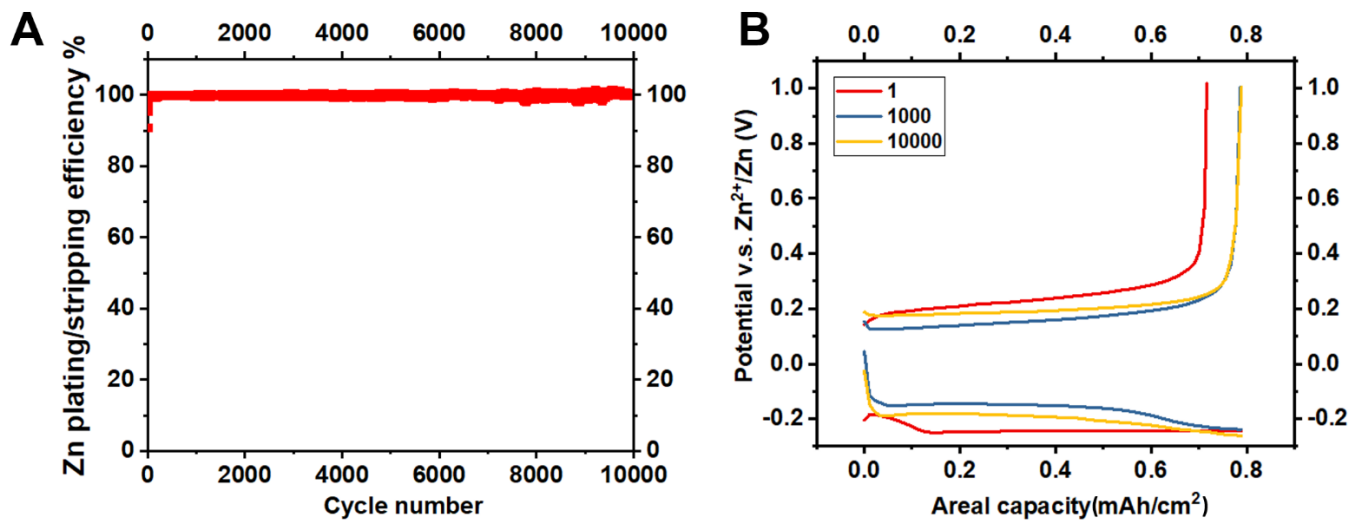

**Supplementary Figure S29. Plating/stripping of Zn on Cu at 40 mA/cm².** (A) CE values and (B) representative plating/stripping voltage profiles. The cell shows an extremely stable plating/stripping behavior over prolonged cycling, with an average CE of 99.94%.

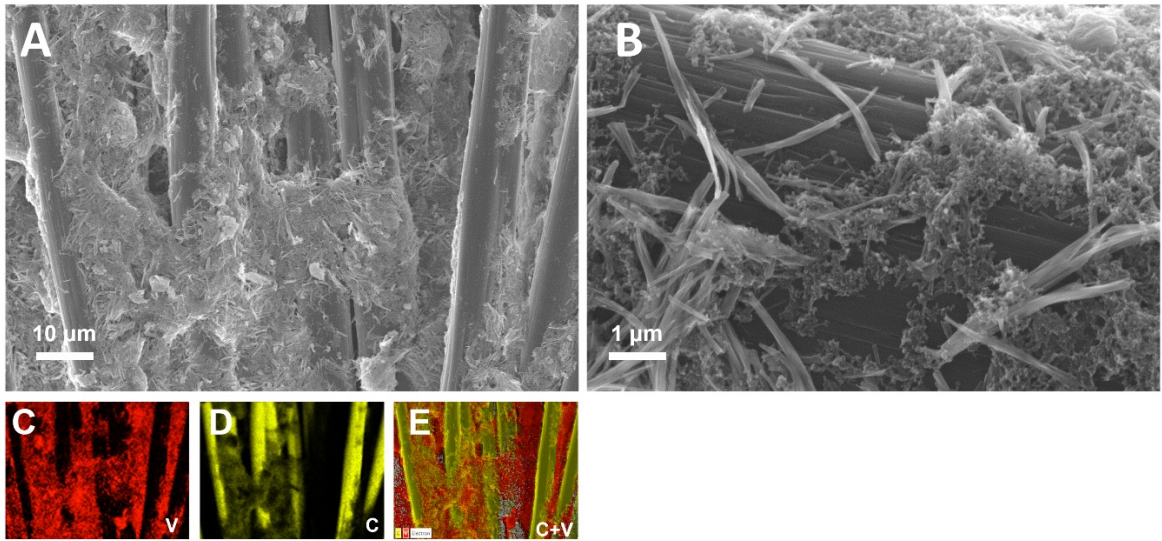

**Supplementary Figure S30. SEM characterization of the sodium vanadate (NVO) cathode.** (A)(B) SEM images. (C)(D)(E) EDS mapping of the region shown in panel (A). The results show that the NVO adopts the nanobelt morphology as reported. The active materials are successfully dispersed into the pores among the interconnected carbon fibers.

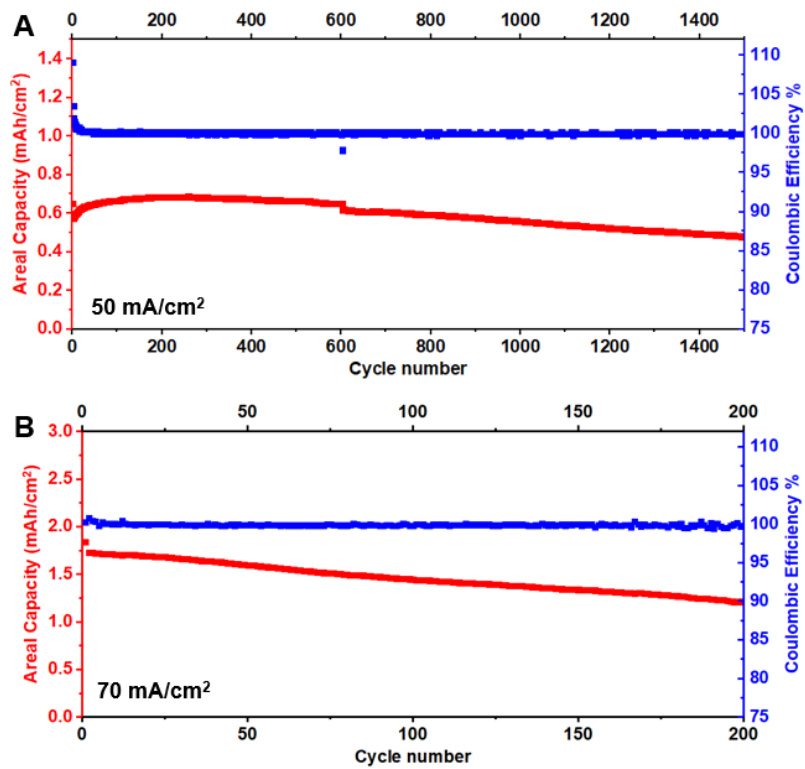

**Supplementary Figure S31. Full battery cycling performance of using "Zn on Cu" anode and NVO cathode.** The N:P ratio is kept at 3:1. The electrode area is ~ 1 cm$^2$.

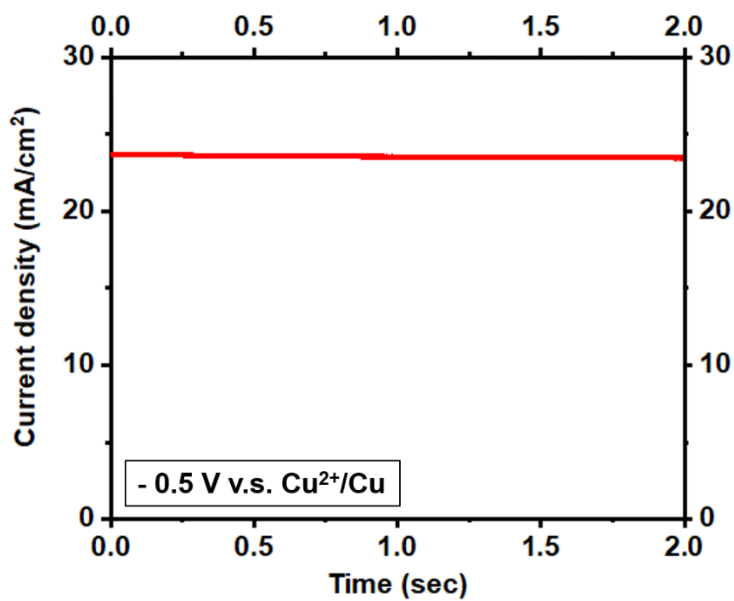

**Supplementary Figure S32. Electrodeposition of Cu onto stainless steel substrate.** The system is held at -0.5 V v.s. $Cu^{2+}/Cu$ for 2 seconds each time, and rest for 1 minute. This process is repeated for 20 times. The high potential is needed to drive uniform deposition, and the rest prevents mass transport limit-induced porous growth.

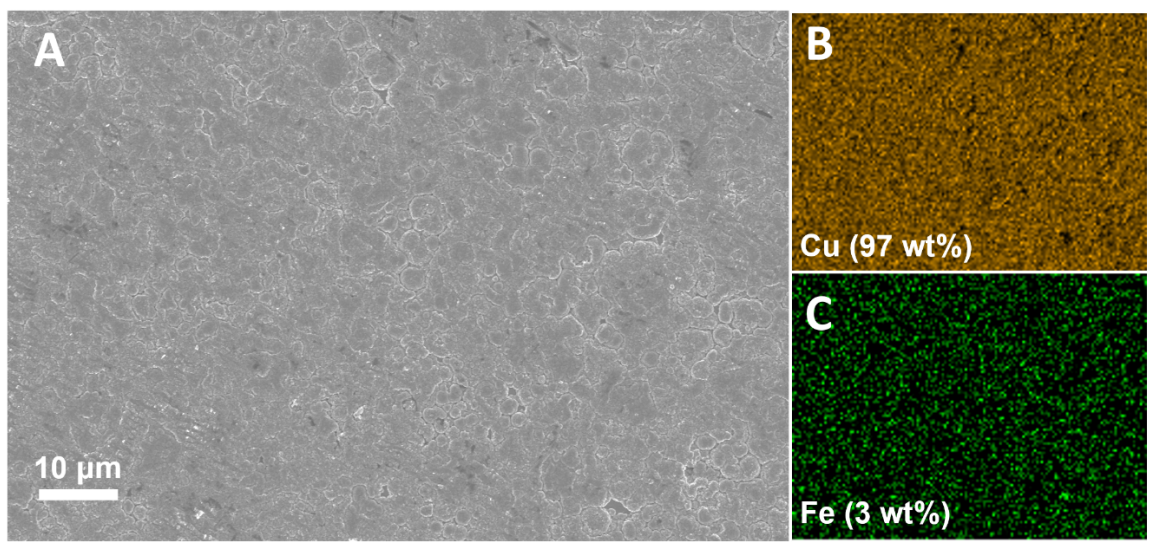

**Supplementary Figure S33. SEM characterization of Cu-coated stainless steel substrate.** (A) SEM image and (B)(C) the corresponding EDS mapping results. The Cu coating shows a high uniformity and surface coverage, as evidenced by both the morphology and the composition.

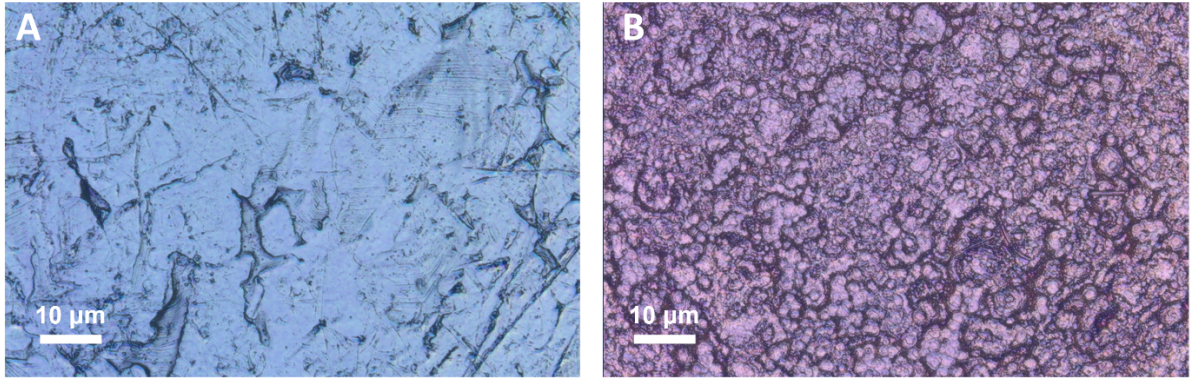

**Supplementary Figure S34. Optical microscopy (OM) characterization of Cu-coated stainless steel substrate.** Images of the stainless steel (A) before and (B) after the Cu deposition. After the deposition, the stainless steel surface is uniformly covered by Cu (see the characteristic color of Cu).

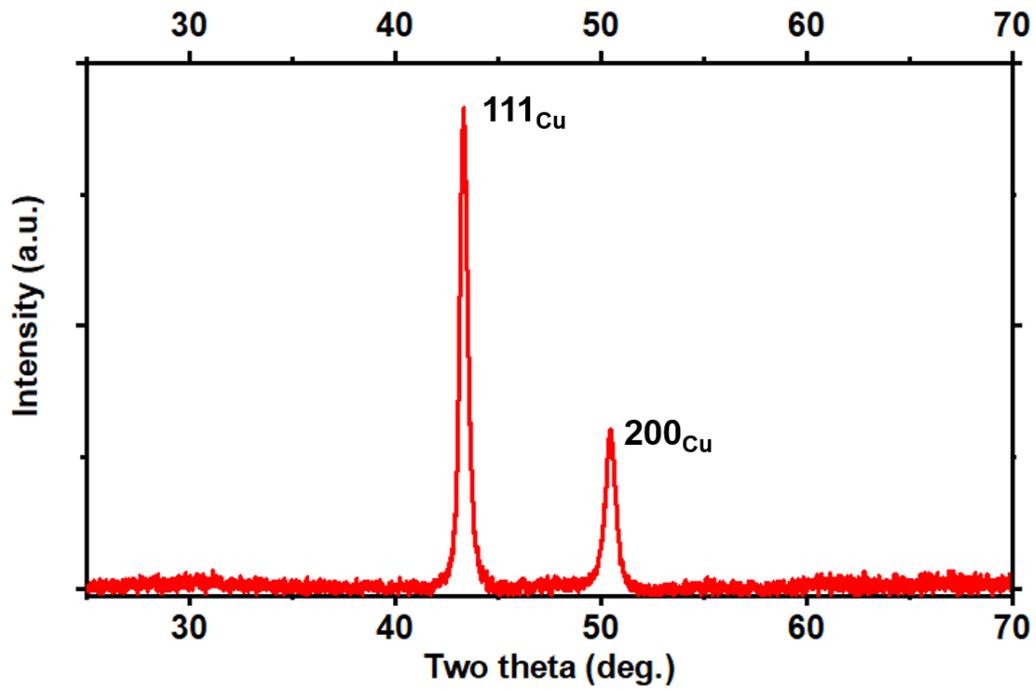

**Supplementary Figure S35. XRD pattern of Cu-coated stainless steel substrate.** The result shows that the surface is fully covered by Cu. This is consistent with the SEM and OM characterization.

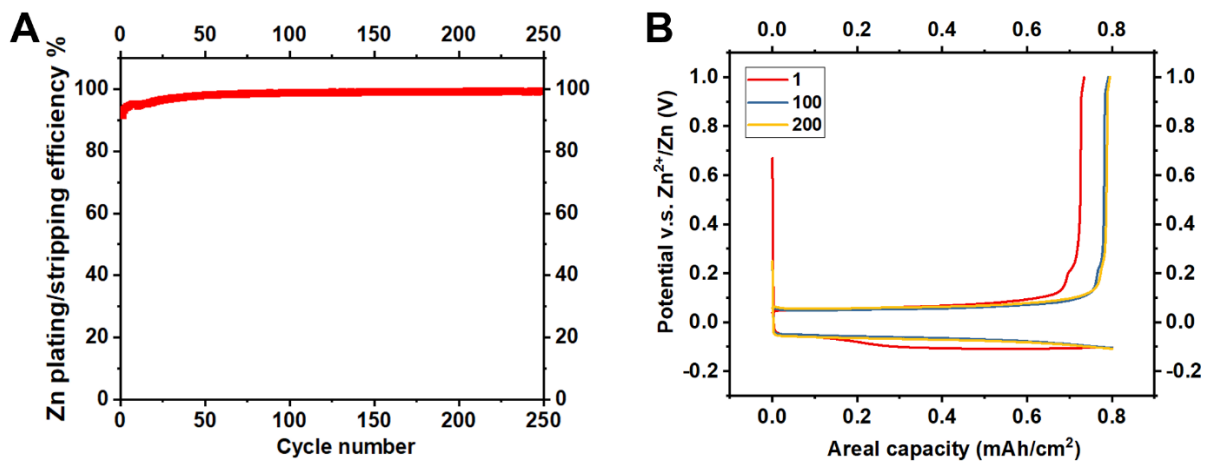

**Supplementary Figure S36. Plating/stripping of Zn on Cu coated SS. Current density: 8 mA/cm².** (A) CE values and (B) representative plating/stripping voltage profiles. No potential evolution suggestive of Zn orphaning is observed during the repeated Zn plating/stripping. The characteristic potential signatures of Cu are detected—see the initial nucleation behavior in the plating branch and the secondary plateau in the stripping branch.

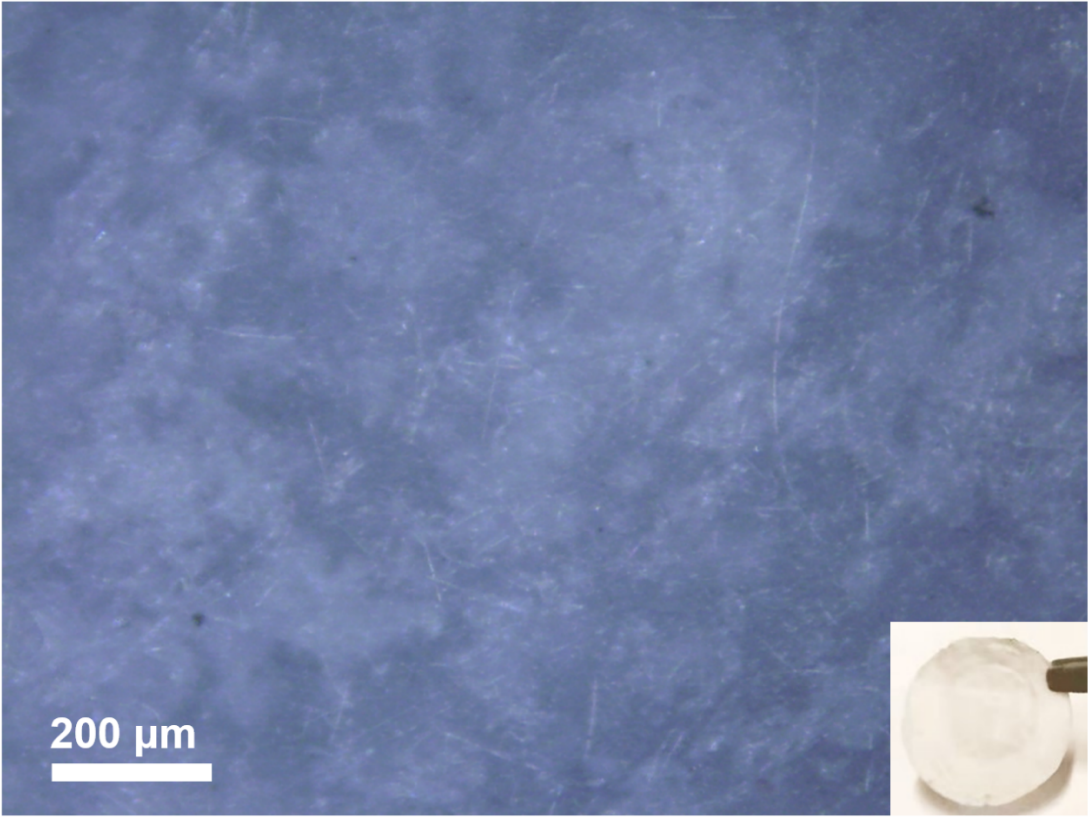

**Supplementary Figure S37. Optical microscopy characterization of the separator facing Cu-coated stainless steel foil after one-time Zn deposition.** The deposition is performed at 8 mA/cm², 0.8 mAh/cm². The result shows that no observable orphaned Zn fragment is formed when the Zn is deposited onto the Cu-coated stainless steel substrate. This observation is consistent with the high plating/stripping efficiency of Zn shown in **Fig. S36**.

**Supplementary Note S1. Evaluation of Na metal plating/stripping.**

We extend the study to sodium as a second metallic anode system for evaluating the proposed concept. This choice is motivated in part by sodium's profound differences with Zn—sodium is soft, highly reactive, and like Li has a low reduction potential. Despite these differences, we report that the "Sabatier"-like principle can be used as effectively to understand reversibility Sodium (Na) metal anodes as for Zn. For brevity, we focus here on result from the Na-Au system for in-depth evaluation. Consistent with our earlier findings, highly reactive alkali metals normally fall into either a non-interacting or a too-strongly-interacting regime with the substrate (*i.e.*, the far-left or the far-right side of the plot shown in **Fig. 5E**). For example, Na-Au shows a -0.283 eV formation enthalpy per atom according to DFT calculation.

As a first assessment of the Na-Au couple, we performed cyclic voltammetry using Au foil as a working electrode. Consistent with the XRD results (**Fig. 7A**) and in analogy to the Zn-Au couple, Na shows an alloying peak on Au substrate at around 0.45 V v.s. $Na^+/Na$ (**Fig. 7B**). Further analysis reveals that the peak current density and the square root of scan rate exhibit a strong linear relation, which confirms the diffusion-controlled nature of the Na-Au alloying reaction occurring above 0 V v.s. $Na^+/Na$ (**Fig. 7C**). Na-Au would therefore be considered in the "too-strongly-interacting" regime. Consistent with this analysis, large secondary capacity plateaus are observed in the Na plating/stripping experiments performed on Au foils.

As a step further, we next considered the challenge of designing an optimal substrate for Na. The fundamental reason that a too strong interaction is unfavorable primarily originates from two potential sources: (**a**) strong alloying of Na in Au causes large structural change and pulverization of Au; and (**b**) strong-alloyed Na-Au leads to incomplete stripping. In response to these two issues, we hypothesize that maintaining the thickness of the Au layer below a certain value could effectively mitigate the unfavorable structural change and the incomplete stripping. It is known that the characteristic timescale ($\tau \approx L^2/D$) for Na diffusion into and out of an Au layer is a strong function of L, the thickness of the layer, and D, the diffusivity. By reducing the Au coating thickness from 1 micron to 100 nm, the characteristic relaxation time τ decreases by as much as two orders of magnitudes! It is likewise known that downsizing is an effective approach for mitigating pulverization of alloying electrodes, with the most widely studied example being silicon electrodes for Li-ion batteries. Multiple independent literature reports have in fact demonstrated that a critical size exists below which the mechanical pulverization is minimized, and the system is instead kept in an elastic regime (62).

The voltage profiles obtained during galvanostatic plating/stripping on various substrates (**Fig. 7D**) provide a straightforward test for these ideas. (**a**) On a stainless-steel substrate, which shows no interaction with Na, there is no alloying capacity above 0 V versus $Na^+$/Na upon deposition, and the stripping features a spiky voltage profile that is suggestive of unstable, incomplete dissolution of Na metal due to the formation of dead Na. This finding is consistent with our earlier detailed study of Na deposition on stainless-steel (63) as well as other metal coatings (64). It is a defining characteristic of cycling a metal anode on the left side of the Sabatier relation shown in **Fig. 5**, *i.e.*, cases where a too-weak interaction causes metal orphaning. (**b**) On a foil-type, bulk Au substrate (~25 micron thick), a significant amount of alloyed Na is observed (approx. 40%). And, less spiky stripping voltage profiles are seen, but the stripping is incomplete as evidenced by the lower Coulombic efficiency. This a characteristic of cycling a metal anode on the right side of the Sabatier curve as shown in **Fig. 5**, i.e., cases where a too-strong interaction causes incomplete stripping. (**c**) On Au coated SS substrates with different Au coating thicknesses, importantly, we find a transition from such bulk conversion (that strands Na) to an optimal regime, wherein the alloying capacity of Na-Au is comparable to what is observed on Zn-Cu. In this case, while the system tends to undergo an unfavorable bulk conversion, the geometry of the thin coatings sets the fundamental limit of the depth that such conversion can reach. Such a transition from unfavorable bulk conversion to a favorable interfacial interaction occurs as Au coating enters the nanoscopic regime, *e.g.*, 100 nm (**Fig. 7D~F**), as evidenced by the >99% plating/stripping reversibility. We hypothesize that this is a critical dimension below which the too-strong interaction between Na-Au is mitigated by the geometric thickness, driving it into an interfacial (as opposed to bulk) interaction regime, as we observed on Zn-Cu. This exemplifies that thickness serves as knob for tuning the nature of such alloying reactions; by carefully designing the dimension of the substrate alloying layer, we could achieve heterointerfacial chemistries of the right strength level for reversible battery metal anodes. This can be compared by analogy to our recent findings related to interfacial Al-O-C bonding producing large enhancements in Al metal anode reversibility in $AlCl_3$-ImCl ionic-liquid electrolyte melts (65).

Taken all together, our observations for the Na-Au system imply: (**a**) The Sabatier-like principle is as valid for Na as it is for Zn. Considering the profound differences in reactivity, mechanics, and electrochemistry of these metals, the principle can be used to define reversibility regimes for heterointerfacial metal plating/stripping processes for any metal. (**b**) The thickness of the substrate provides a tunable variable that controls the nature of the alloying reactions that intrinsically fall onto the right side of the Sabatier relation. It sets a geometric limit on the alloying reaction, preventing it from entering bulk conversion/pulverization. (**c**) The results of Na metal are of immediate interest to the Na battery community, and together with the Zn results, are important to the development of metal anodes including Li, Na, K, Ca, Mg, Zn, Al, *etc*.